\documentclass[a4paper,11pt]{article}
\usepackage{jheppub}

\usepackage[utf8]{inputenc}
\usepackage{amsmath}
\usepackage{amssymb}
\usepackage{graphicx}
\usepackage{multirow}
\usepackage[normalem]{ulem}

\newcommand{\orcid}[1]{\href{https://orcid.org/#1}{#1}}
\newcommand{\e}[1]{\times10^{#1}}
\newcommand{\eps}{\epsilon}

\title{Testing New Physics in Oscillations at a Neutrino Factory}

\author[1]{Peter B.~Denton\note{\orcid{0000-0002-5209-872X}},}
\author[2]{Julia Gehrlein\note{\orcid{0000-0002-1235-0505}},}
\author[3,4]{and Chui-Fan Kong\note{\orcid{0009-0007-7010-5085}}}

\affiliation[1]{High Energy Theory Group, Physics Department, Brookhaven National Laboratory, Upton, NY 11973, USA}
\affiliation[2]{Physics Department, Colorado State University, Fort Collins, CO 80523, USA}
\affiliation[3]{Tsung-Dao Lee Institute \& School of Physics and Astronomy, Shanghai Jiao Tong University, China}
\affiliation[4]{Key Laboratory for Particle Astrophysics and Cosmology (MOE) \& Shanghai Key Laboratory for Particle Physics and Cosmology, Shanghai Jiao Tong University, Shanghai 200240, China}

\emailAdd{pdenton@bnl.gov}
\emailAdd{julia.gehrlein@colostate.edu}
\emailAdd{kongcf@sjtu.edu.cn}

\makeatletter
\hypersetup{colorlinks=true,allcolors=[rgb]{1,0.56,0},pdftitle=\@title}
\makeatother

\abstract{
A neutrino factory is a potential successor to the upcoming generation of neutrino oscillation experiments and a possible
precursor to next-generation muon colliders.
Such a machine would provide a well-characterized beam of $\nu_\mu$, $\bar\nu_\mu$, $\nu_e$, and $\bar\nu_e$ neutrinos with comparable statistics.
Here we show the sensitivity of a neutrino factory to new oscillation physics scenarios such as vector neutrino non-standard interactions and CPT violation. We study two different potential setups for a neutrino factory with different assumptions on charge identification in the far detector.
We find that 10 years of a neutrino factory combined with 10 years of DUNE can improve over most of the current constraints on these scenarios and even over forecasted constraints by 20 years of DUNE.
Additionally, we find that a neutrino factory can break degeneracies between the standard oscillation parameters and neutrino non-standard interaction parameters present at DUNE.}

\begin{document}

\maketitle

\section{Introduction}
Neutrino physics has entered the precision era after a non-zero $\theta_{13}$ mixing angle was measured \cite{DayaBay:2012fng,RENO:2012mkc,DoubleChooz:2014kuw,RENO:2018dro,DoubleChooz:2019qbj,DayaBay:2022orm}.
Still, there are three main unknowns in the standard three-flavor neutrino oscillation
framework: the neutrino mass ordering, the octant of $\theta_{23}$, and the value of $\delta_{CP}$ \cite{Denton:2022een,Denton:2025jkt}. 
The next-generation neutrino oscillation experiments like JUNO \cite{JUNO:2015zny}, Hyper-Kamiokande \cite{Hyper-Kamiokande:2018ofw}, and DUNE \cite{DUNE:2020ypp}, are expected to address these unknowns.
However, the experimental sensitivities to the standard oscillation parameters can be reduced in the presence of physics beyond the Standard Model (BSM), such as non-standard interaction \cite{Farzan:2017xzy,Proceedings:2019qno} and CPT violation \cite{Arguelles:2022tki}.
In addition, given the open theoretical and experimental questions related to neutrinos, it is important to search for additional new physics in the neutrino sector wherever possible.

However, there exist degeneracies between many of the new physics parameters and the standard oscillation parameters, which may make it challenging to search for new physics even with the upcoming planned experiments.
There are several experimental proposals beyond the upcoming JUNO, Hyper-Kamiokande, and DUNE programs that aim to break these degeneracies and improve the sensitivities to both BSM parameters and the standard oscillation parameters. 
First, a multi-baseline setup such as the proposed long-baseline experiments T2HKK \cite{Hyper-Kamiokande:2016srs} and ESSnuSB \cite{ESSnuSB:2021azq} targeting the second maximum could provide a different handle on the oscillation physics, as could the proposed P2O experiment \cite{Akindinov:2019flp}.
Second, an additional low-energy neutrino source from muon decay at rest \cite{Ge:2016dlx} can reduce the new physics impact and provide a clean measurement of standard oscillation parameters, which in turn gives a better sensitivity to new physics parameters when combined with a high-energy neutrino beam.
Finally, performing a joint analysis of all experiments is desirable to resolve degeneracies and benefit from complementarities to provide stringent constraints on BSM scenarios (as done in \cite{Coloma:2023ixt} in the case of new interactions).

In this study, we consider another possibility for a potential future neutrino oscillation experiment, namely a neutrino factory (NF) which uses LAr far detectors to measure high-energy $\mathcal{O}$(GeV) $\overset{\scriptscriptstyle(-)}{\nu}_\mu$ and $\overset{\scriptscriptstyle(-)}{\nu}_e$ beams coming from muon decays \cite{Geer:1997iz,DeRujula:1998umv,Blondel:2000gj,Albright:2000xi,Blondel:2006su,FernandezMartinez:2010zza}.
Recently, standard three-flavor oscillations at a modern neutrino factory setup were studied in \cite{Denton:2024glz} as well as non-oscillation new physics in \cite{Bogacz:2022xsj}.
In this paper we study new physics in oscillations at a neutrino factory. We 
focus on vector neutrino non-standard 
interaction (NSI) and CPT violation where long-baseline experiments can provide the strongest bounds for many of the oscillation parameters.
Other BSM scenarios, like scalar NSI \cite{Ge:2018uhz,Babu:2019iml} are much better 
constrained by solar neutrino experiments due to the significant amplification of this effect inside the sun \cite{Denton:2024upc} and the sterile neutrino scenario is typically best constrained by short-baseline experiments \cite{Machado:2019oxb,Dentler:2018sju,Boser:2019rta} and atmospheric neutrinos \cite{Nunokawa:2003ep,IceCubeCollaboration:2024nle}.
A neutrino factory mainly benefits from three perspectives to constrain NSI and CPT violation compared to existing and planned oscillation experiments, like DUNE or Hyper-Kamiokande, which use neutrinos produced in a fixed target setup: i) a larger statistics of the oscillation channels from the electron flavor to the electron and muon flavors makes it sensitive to more new physics parameters, ii) the higher neutrino energy can amplify the new physics effects such as NSI allowing for an easier observation of their effects, and iii) the simultaneous presence of equally large amounts of neutrinos and anti-neutrinos, albeit with a different energy spectrum, in the beam allows for studies of new physics in the neutrino and anti-neutrino sectors separately.
Vector NSIs have been studied in the context of a NF or similar experiments in the past, see e.g.~\cite{Huber:2001zw,Huber:2002bi,Apollonio:2002en,Blennow:2005qj,ISSPhysicsWorkingGroup:2007lul,Kopp:2007mi,Kopp:2008ds,Meloni:2009cg,Antusch:2009pm,Bakhti:2016prn} and CPT violation at a neutrino factory has been studied in \cite{Bilenky:2001ka,Antusch:2008zj}; this study makes use of an updated perspective on the oscillation parameters, the global constraints on new physics, and an up-to-date picture of the upcoming experimental landscape.

The paper is outlined as follows.
In section \ref{sec:NF}, we discuss the experimental setup of a neutrino factory.
In sections \ref{sec:nsi} and \ref{sec:cpt} we present our results for the the scenarios of new neutrino interactions and CPT violation.
We conclude in sec.~\ref{sec:conclusions}.

\section{Experimental Setup}
\label{sec:NF}
At a neutrino factory an intense neutrino beam with $\mathcal{O}$(GeV) energy
can be produced via the decays of positively 
(negatively) charged muons inside a muon storage ring. 
Due to the 3-body muon decay process,
the flavor composition and energy spectrum 
of the neutrinos are well understood. Moreover, 
different from traditional 
accelerator neutrino experiments, there exists a large 
fraction of $\nu_e$ ($\bar\nu_e$) from $\mu^+$ ($\mu^-$) 
decay which allows for studies 
of $\nu_e\rightarrow \nu_e,\nu_\mu$ ($\bar\nu_e\rightarrow \bar\nu_e, \bar\nu_\mu$) oscillation channels with large statistics.
We consider two possible setups for a neutrino factory, motivated by existing accelerator complexes in the US.\footnote{See \cite{Kitano:2024kdv} for a study of a neutrino factory based at J-PARC.} Either the neutrinos are produced 
at i) Fermilab (FNAL) or at ii) Brookhaven National Laboratory (BNL) \cite{Denton:2024glz}. We assume $10^{21}$ muon decays per year \cite{ISSDetectorWorkingGroup:2007uvu,Geer:2010zz} with equal running time for muons and anti-muons. For the FNAL-SURF (BNL-SURF) baseline we use as maximal neutrino energy $E_\mu=5 (8)$ GeV. The choice of energy and running time is motivated by previous results from \cite{Denton:2024glz} as these values optimize the precision of the CP phase $\delta$ in the absence of new physics at a neutrino factory; for different goals such as new physics one may prefer different muon energies. 
We assume that the far detector is the DUNE far detector with a total of 40\,kt LAr fiducial mass \cite{DUNE:2020lwj} to detect neutrinos produced at FNAL or BNL, which have traveled a baseline of $L=1284.9\,$km or $L=2542.3 \,$km, respectively. 
For both configurations we use a matter density of $2.848\,$g/cm$^3$ and we assume a 2\% uncertainty on matter density which is consistent with the DUNE TDR \cite{DUNE:2021cuw}.
Realistically, the matter density for the BNL configuration is somewhat higher due to its longer baseline however, similarly to DUNE, a neutrino factory is expected to be sensitive to the matter density at the 30\% level \cite{Kelly:2018kmb}, thus a slightly different value due to the longer baseline will not significantly affect our results.

The beam at a neutrino factory has a large intrinsic background, for example in the $\nu_e$ disappearance channel coming from $\bar\nu_\mu\to \bar\nu_e$
appearance.
It is possible to differentiate the two channels at some level due to the different initial spectrum combined with the (presumably) different oscillation probability.
The excellent energy reconstruction abilities of LArTPC detectors can partially separate these components at the far detector \cite{Huber:2008yx} and additionally
a detector with charge identification (CID) capabilities allows for improved background rejection. 
To study the role of CID we consider three scenarios of the detector performance: i) no charge identification, ii) with 100\% charge identification of 
electrons ($e$CID), and iii) with 100\% charge identification of muons ($\mu$CID).
Existing studies have proposed a variety of possible CID scenarios for the DUNE far detectors.
For example, for muon neutrino CID, DUNE could differentiate a muon final state from an anti-muon final state with $\epsilon_{CID}\approx 72\%$ \cite{Ternes:2019sak} from muon capture on argon.
In addition, DUNE may get some CID information from inelasticity as well.
In \cite{Huber:2008yx} it has been pointed out that CID can also be achieved statistically at some level, provided that the energy resolution is sufficiently good.
Electron CID at GeV energies has been studied in the context of a magnetized LAr detector \cite{Rubbia:1977zz,Rubbia:2001pk,Rubbia:2004tz,Rubbia:2009md}.
Other detector configurations may achieve significantly improved CID capabilities; given that the final detector details for DUNE are still under consideration, such a discussion is extremely timely.

We also include the oscillated tau neutrinos with the charged-current detection process at the far detector.
Due to the large uncertainties on the tau neutrino signal efficiency at LAr detectors, however, we treat the tau neutrino events as background only, using the cross section from \cite{Yaeggy:2023wfe}.
We consider only normalization uncertainties at the far detector for a neutrino factory where we use 2.5\% for both the muon and electron neutrino normalization uncertainties because for a neutrino factory the statistics of $\nu_e,~\nu_\mu$ in the FD are comparable.
This differs from the approach used for LBNF neutrinos \cite{DUNE:2021cuw}, where the muon and electron neutrino signal
normalization uncertainties are taken to be 5\% and 2\%, respectively, while a common 5\% uncertainty is used for muon and electron neutrino background as the muon neutrino measurement is used to inform the electron neutrino flux.
We also take a 10\% uncertainty on the neutral-current background and a 20\% uncertainty on the tau background.
We use the GLoBES framework \cite{Huber:2004ka, Huber:2007ji} for our studies. 

To compare the potential of a NF to planned future experiments, we also take DUNE
into account\footnote{We note that there has been a high energy tune for DUNE proposed designed to enhance its sensitivity for tau neutrinos, NSIs, and other new physics scenarios \cite{Masud:2017bcf,Masud:2018pig,DeGouvea:2019kea,Rout:2020cxi,Ghoshal:2019pab,Siyeon:2024pte}. However this advantage comes at the cost of shifting away from the CP optimized beam.
The NF beam parameters we have chosen here were also selected to maximize the sensitivity to CP violation \cite{Denton:2024glz}.
Thus for a direct comparison, we focus on DUNE in the nominal configuration.}, where the experimental 
details can be found in \cite{DUNE:2020ypp,DUNE:2021cuw}. 
We consider the following experimental setups: 
 i) 20 years of (DUNE-20)\footnote{We assume $1.1\times 10^{21}$ POT per year of DUNE \cite{DUNE:2021cuw}.}, or 
ii) 10 years of DUNE plus 10 years of a neutrino factory at either FNAL (NF-FNAL) and BNL (NF-BNL) baseline.
Each of the two NF scenarios has three CID scenarios as mentioned above.
Therefore we have seven different experimental setups in total. To keep the plots for the results clean and to limit the number of their curves, we show the results for the BML configuration in the appendix.

\begin{table}
    \centering
    \caption{Overview of the neutrino oscillation parameters and matter density used in this study.
    The oscillation parameters come from the global fit \cite{Esteban:2020cvm}.
    We assume the current best fit values as the true values for these oscillation parameters.}

    \begin{tabular}{c|c|c}
        & \textbf{Best Fit} & \textbf{1$\sigma$ Range} \\ 
        \hline
        $\sin^2 \theta_{12}$ 
        & 0.307 
        & Fixed
    \\ 
        $\sin^2 \theta_{23}$ 
        & 0.572
        & Free
    \\ 
        $\sin^2 \theta_{13}$ 
        & 0.02203
        & $\pm 5.8\times 10^{-4}$
    \\ 
        $\delta_{\text{CP}}$ 
        & $-90^\circ$
        & Free
    \\ 
        $\Delta m^2_{21}~[10^{-5} {\rm eV}^2]$ 
        & 7.41
        & Fixed
    \\ 
        $\Delta m^2_{31}~[10^{-3} {\rm eV}^2]$ 
        & 2.511
        &  Free
    \\ 
        Matter density [g/cc]
        & 2.848
        & 2\%
    \end{tabular}
\label{tab:oscillation-params}
\end{table}

We summarize the assumed true values for the standard oscillation parameters in table \ref{tab:oscillation-params}, taken from \cite{Esteban:2020cvm}.\footnote{Different \cite{deSalas:2020pgw,Capozzi:2021fjo} or more} updated global fit results \cite{Esteban:2024eli} do not significantly change our results as either DUNE or a NF will measure most parameters relevant for new physics searches better than the existing constraints.
We assume true normal ordering throughout this study.
For $\delta_{CP}$ 
we assume its true value corresponds to the largest allowed CP violation given the existing measurements, $\delta_{CP}=-90^\circ$, unless otherwise stated.
This value is also somewhat preferred by T2K \cite{T2K:2023smv}, but we assume no prior on $\delta_{CP}$.
Note that the preferred value is in minor tension with the 
NO$\nu$A result \cite{NOvA:2021nfi,2852068}, where the best fit is 
around $150^\circ$ for normal ordering and $-90^\circ$ for 
inverted ordering.
We fix the solar mixing angle $\theta_{12}$ and mass-squared difference $\Delta m^2_{21}$ to their current best fit values from \cite{Esteban:2020cvm}
as long-baseline accelerator experiments have limited sensitivity to them \cite{Denton:2023zwa}. Since the reactor mixing angle 
$\theta_{13}$ has been
measured very precisely by reactor experiments \cite{DayaBay:2022orm,RENO:2018dro,DoubleChooz:2019qbj}, 
we fix its best fit value. 
We take $\theta_{23}$ and $\Delta m^{2}_{31}$ 
parameters as free parameters without any prior 
since they will be well measured at long-baseline experiments.

\section{Neutrino Non-Standard Interactions}
\label{sec:nsi}
We first introduce the framework for vector NSI and then show the sensitivity of DUNE and a NF to the NSI parameters and the standard oscillation parameters in the presence of NSI. We focus on $\delta_{CP}^\text{true}=-90^\circ$, the results for $\delta_{CP}^\text{true}=0^\circ$ can be found in the appendix.

\subsection{Framework}
Neutrinos could have interactions with other fermions due to the presence of new mediators. A general framework to incorporate this effect is neutrino non-standard interactions which, for example, originate from dimension-6 operators.
First proposed by Wolfenstein \cite{Wolfenstein:1977ue}, NSIs are often considered to have a vector mediator.
In fact, such new neutrino interactions are natural features in many neutrino mass models with new mediators \cite{Ohlsson:2012kf,Miranda:2015dra,Babu:2019mfe}. 
The strength of the NSI depends on the specific model and large NSI is predicted in some models, see e.g.~\cite{Forero:2016ghr,Denton:2018dqq,Dey:2018yht,Babu:2017olk,Farzan:2016wym,Farzan:2015hkd,Farzan:2015doa,Babu:2019mfe}.
In general, vector NSI can be of charged-current (CC) or neutral-current (NC) type \cite{Farzan:2017xzy,Proceedings:2019qno}.
Both affect neutrinos during their propagation through matter while only CC NSI affects neutrinos at the point of production and detection.
Note also that NC NSI does affect the detection of neutrinos in NC channels, see e.g.~\cite{Coloma:2017egw,Gehrlein:2024vwz}.
Long-baseline experiments like DUNE and a NF are strongly affected by matter effects, and thus are expected to have exceptional sensitivity to the NC NSI.
To this end, we take the NC NSI into consideration below in the context of propagation.\footnote{One should also consider NC NSI in the detection of the NC channel at a NF. We leave such an analysis at a NF for future work.}

In the low-energy regime, neutrino 
NC NSI with matter fields can be formulated in terms of the effective four-fermion Lagrangian term \cite{Wolfenstein:1977ue}, (see \cite{Antusch:2008tz,Biggio:2009nt,Ohlsson:2012kf,Miranda:2015dra,Farzan:2017xzy,Proceedings:2019qno} for recent reviews),
\begin{subequations}
\begin{align}
  \mathcal{L}_{\rm NSI}
= 
  -2\sqrt{2} G_F \sum_{f, \alpha, \beta}
  \epsilon_{\alpha \beta}^{f, V} (\bar\nu_\alpha \gamma^\mu P_L \nu_\beta)
  (\bar f \gamma_\mu P f)\,,
\end{align}
\end{subequations}
where $G_F$ is the Fermi constant.
The dimensionless coefficient $\epsilon^{f, V}_{\alpha \beta}$ 
quantifies the size of the neutrino NSI 
with matter fermions $f\in \left\{e, u, d\right\}$ relative to the standard weak interaction.
The above effective Lagrangian introduces a new neutrino matter potential which modifies the Hamiltonian,
\begin{equation}
  H
=
  \frac{1}{2E}
  \left[
  U
  \begin{pmatrix}
  0 & 0 & 0\\
  0 & \Delta m^2_{21} & 0 \\
  0 & 0 & \Delta m^2_{31}
  \end{pmatrix}
  U^\dagger
  +
  a
  \begin{pmatrix}
   1+\epsilon_{ee} & \epsilon_{e\mu} & \epsilon_{e\tau}\\
   \epsilon_{e\mu}^\ast & \epsilon_{\mu\mu}
   & \epsilon_{\mu\tau}\\
   \epsilon_{e\tau}^\ast & \epsilon_{\mu\tau}^\ast & \epsilon_{\tau\tau}
  \end{pmatrix}
  \right],
\end{equation}
where $U$ is the PMNS mixing matrix \cite{Pontecorvo:1957qd,Maki:1962mu} and $\Delta m^2_{ij} \equiv m_i^2 - m_j^2$ are the mass squared differences.
Here the first term is the vacuum component $H_{\rm vac}$, while
$a\equiv 2 \sqrt{2} G_F N_e E$
with $N_e$ being the electron number density in the second term
describes the strength of the matter effect.
Finally, we relate the Lagrangian level NSI parameters to the Hamiltonian level ones via $\epsilon_{\alpha\beta} \equiv \sum_{f}N_f \epsilon_{\alpha\beta}^{f,V}/N_e$ with the fermion matter density $N_f$. 

The diagonal NSI parameters are real, while the off-diagonal NSI parameters may be complex and can be parametrized as,
\begin{equation}
  \epsilon_{\alpha \beta}
= 
  |\epsilon_{\alpha\beta}| e^{\text{i}\phi_{\alpha\beta}}.
\end{equation}
Therefore, there are a total of nine new real NSI parameters.
However, one of the diagonal parameters
can be subtracted without affecting the 
oscillation behavior.
For definiteness, we
adopt the notation, $a\times{\rm Diag}(1+\epsilon_{ee}-\epsilon_{\mu\mu},0, \epsilon_{\tau\tau}-\epsilon_{\mu\mu})$, for the diagonal part of the matter potential in our study.

\subsection{Results for non-standard vector interactions}
\begin{figure}
\centering
\includegraphics[height=0.37\textwidth]{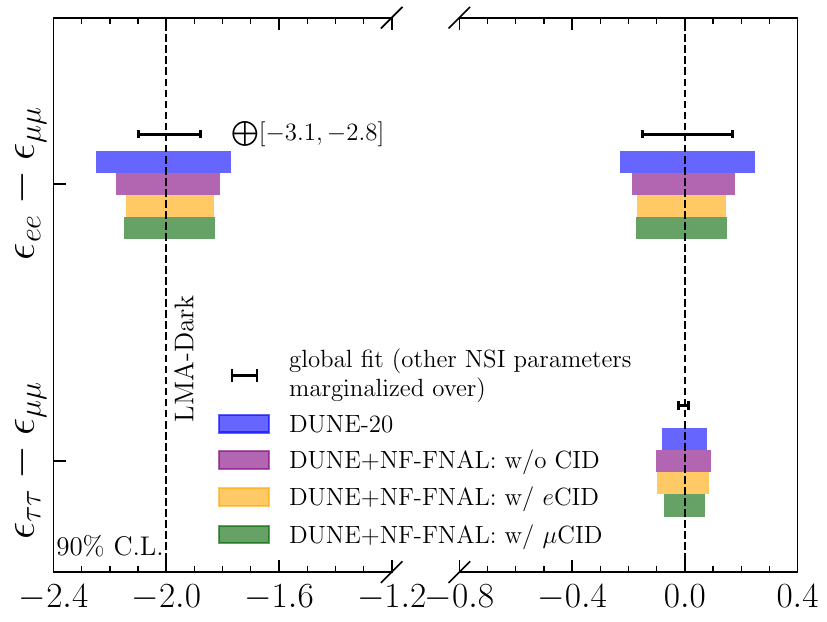}
\includegraphics[height=0.37\textwidth]{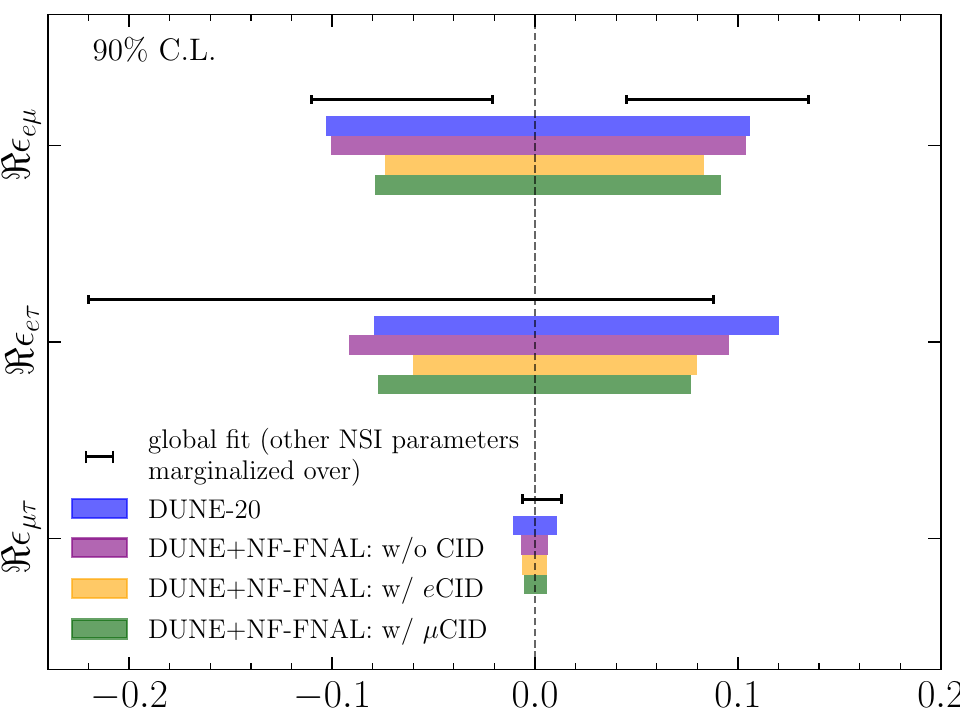}
\caption{Constraints on the diagonal (left) NSI parameters and off-diagonal NSI parameters (taken to be real) (right) at 90\% confidence level assume 20 years of DUNE (blue) and 10 years of DUNE combined with 10 years of a NF located at FNAL. The purple, yellow, and green color correspond to different levels of CID at the NF far detector: no CID, 100\% electron CID, and 100\% $\mu$ CID, respectively. We compare the results to the global fit results from \cite{Coloma:2023ixt}. There are three allowed regions of $\epsilon_{ee}-\epsilon_{\mu\mu}$ from the global fit, where two of them are shown by error bars while the remaining one $[-3.1, -2.8]$ is shown with numbers.
See Fig.~\ref{fig:NSI 1D BNL} in the appendix for the sensitivities of the BNL configuration.}
\label{fig:NSI 1D}
\end{figure}

We start with the simplest case, where all NSI parameters are assumed to be real and only one NSI parameter is considered at a time; we will generalize this assumption later.
While this will miss some of the more interesting physics cases presented below, it is valuable in highlighting the relative role of each experiment.
In Fig.~\ref{fig:NSI 1D}, we show the expected constraints on the diagonal and the real part of the off-diagonal NSI parameters at the 90\% confidence level (C.L.) in the left and right panels, respectively.
We focus on four of the experimental scenarios discussed in Sec.~\ref{sec:NF}.
The allowed regions are shown by patches of different colors.
For comparison, we also include the current global fit data 
\cite{Coloma:2023ixt} with an error bar. Note however that for the global fit results the NSI parameters not shown have been marginalized over.

We find that for all five NSI parameters,
DUNE combined with a neutrino factory can reach
better sensitivities than DUNE alone, even with doubled running time.
As we show in the appendix, especially for a neutrino factory at BNL with $e$/$\mu$CID, the sensitivity can 
be improved by a factor of $\sim2$ compared to DUNE with 20 years running time.
We also generally find that it is possible to improve beyond global bounds even in cases where non-accelerator experiments are expected to dominate.
The first of which is $\eps_{ee}-\eps_{\mu\mu}$ which is strongly constrained by the combination of solar and reactor data, and yet DUNE and a NF may be better than existing constraints.
The second of which are the two parameters tightly constrained in atmospheric experiments: $\eps_{\tau\tau}-\eps_{\mu\mu}$ and $\eps_{\mu\tau}$.
We find that DUNE and a NF can actually outperform the existing constraints on $\eps_{\mu\tau}$ from a recent global fit \cite{Coloma:2023ixt} as well as newer similar constraints directly from the experiments, see \cite{KM3NeT:2024pte}.
Finally, we note that in terms of the NSI parameters expected to be best probed by long-baseline accelerator experiments, $\eps_{e\mu}$ and $\eps_{e\tau}$, we find that DUNE with a NF can outperform, in some cases easily, existing constraints.

There are two separate 
regions in the $\epsilon_{ee}-\epsilon_{\mu\mu}$ parameter space due to the LMA-Dark solution \cite{Miranda:2004nb,Escrihuela:2009up,Bakhti:2014pva,Coloma:2016gei,Denton:2021vtf,Denton:2022nol}. 
This is an exact degeneracy in the oscillation Hamiltonian in matter in the presence of NSI which requires scattering experiments to break it, see e.g.~\cite{Coloma:2017egw,Coloma:2017ncl,Chaves:2021pey,Denton:2022nol}.
The Hamiltonian in the case of the standard 
matter effect is physically equivalent to the 
Hamiltonian $H$ in the presence of NSI
by taking $\Delta m^2_{21}\to -\Delta m^2_{21}$, $\Delta m^2_{31}\to-\Delta m^2_{31}$, $\delta_{CP}\to -\delta_{CP},$ $\epsilon_{ee}-\epsilon_{\mu\mu}=-2$, $\eps_{\tau\tau}-\eps_{\mu\mu}=0$, and $\eps_{\alpha\beta}\in\mathbb R$.

\begin{figure}
\centering
\includegraphics[width=1\textwidth]{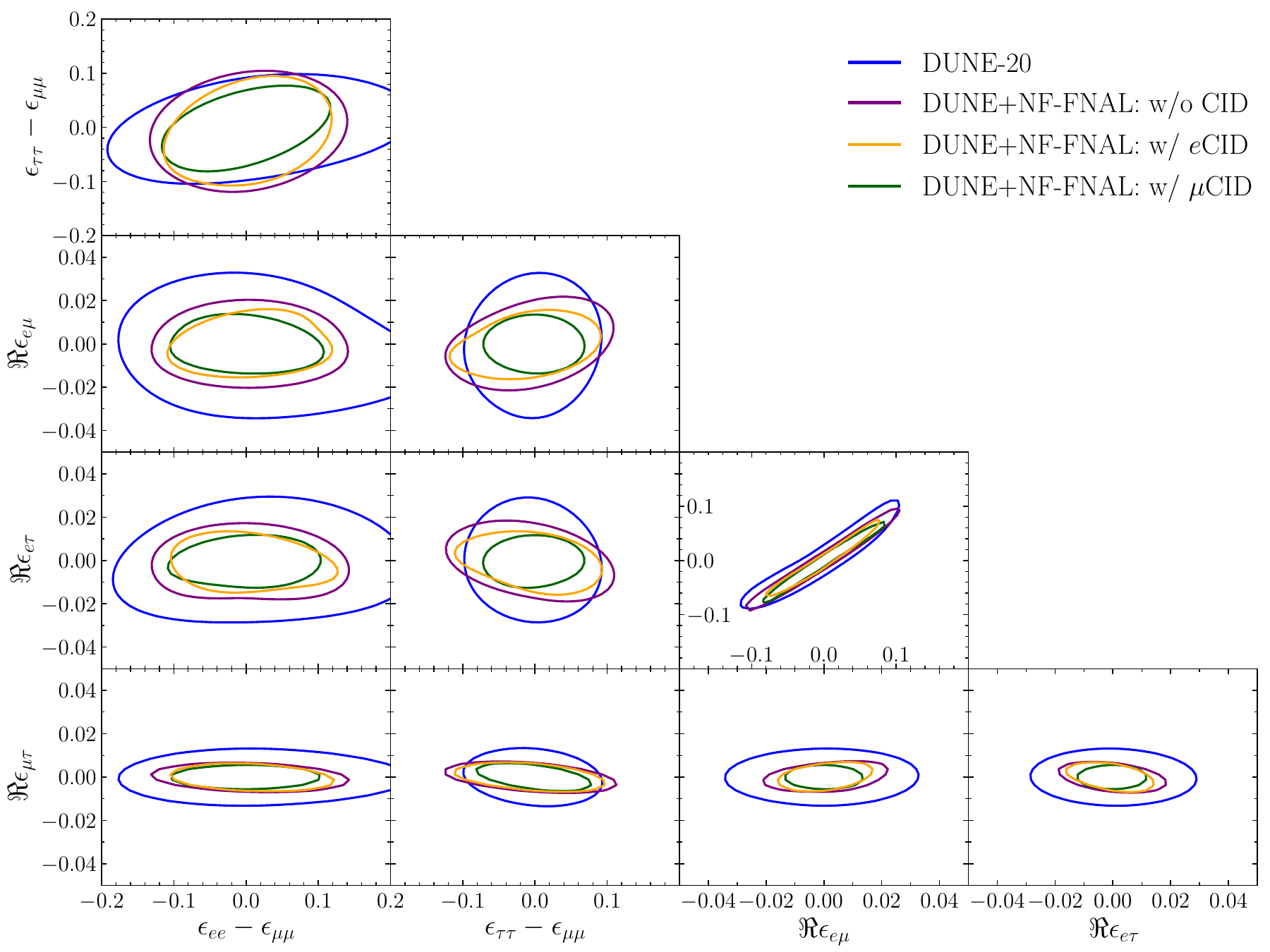}
\caption{Expected constraints on 
different combinations of two non-zero NSI parameters
at 90\% confidence level. 
All NSI parameters are assumed to be real, that is, the complex phase is $\in \{0,\pi\}$.
We compare the expected constraints from 20 years of DUNE only and 10 years of DUNE combined with 10 years of NF-FNAL for different assumptions on CID.
Note that the $\eps_{e\mu}$-$\eps_{e\tau}$ panel has different axes.
See Fig.~\ref{fig:NSI two-D BNL} in the appendix for the sensitivities for the BNL configuration.}
\label{fig:NSI two-D}
\end{figure}

To study other partial degeneracies among different NSI parameters, we extend our study to the two-parameter space which we anticipate should cover the majority of the degeneracy space.
We first consider the case where the off-diagonal NSI parameters are assumed to be real and the experimental constraints at 90\% C.L.~are shown in 
Fig.~\ref{fig:NSI two-D}. There exists a significant
degeneracy between $\Re \epsilon_{e\mu}$ and $\Re \epsilon_{e\tau}$, see e.g.~\cite{Liao:2016hsa,Chatterjee:2021wac}, which is because the dominant combinations of NSI parameters in the $\nu_e/\bar\nu_e$ appearance channel are $\epsilon_{e\mu}+\epsilon_{e\tau}$ and $\epsilon_{e\mu}-\epsilon_{e\tau}$, assuming real NSI parameters \cite{Denton:2022pxt}. Especially when $\delta_{CP}=-90^\circ$, only the combination of $\epsilon_{e\mu}-\epsilon_{e\tau}$ survives \cite{Denton:2022pxt}, which leads to the positive correlation between $\epsilon_{e\mu}$ and $\epsilon_{e\tau}$ as shown in the figure.
As we show in the appendix, in all cases except the $\epsilon_{\tau\tau}-\epsilon_{\mu\mu}$-$\epsilon_{ee}-\epsilon_{\mu\mu}$ panel, DUNE combined with a neutrino factory at 
BNL with $\mu$CID configuration gives the most stringent constraints on the two-parameter space due to the longer baseline of the BNL setup.
In the $\epsilon_{\tau\tau}-\epsilon_{\mu\mu}$-$\epsilon_{ee}-\epsilon_{\mu\mu}$ panel, the $e$CID and $\mu$CID cases are both very similar and both provide strong constraints.

\begin{figure}
\centering
\includegraphics[width=\textwidth]{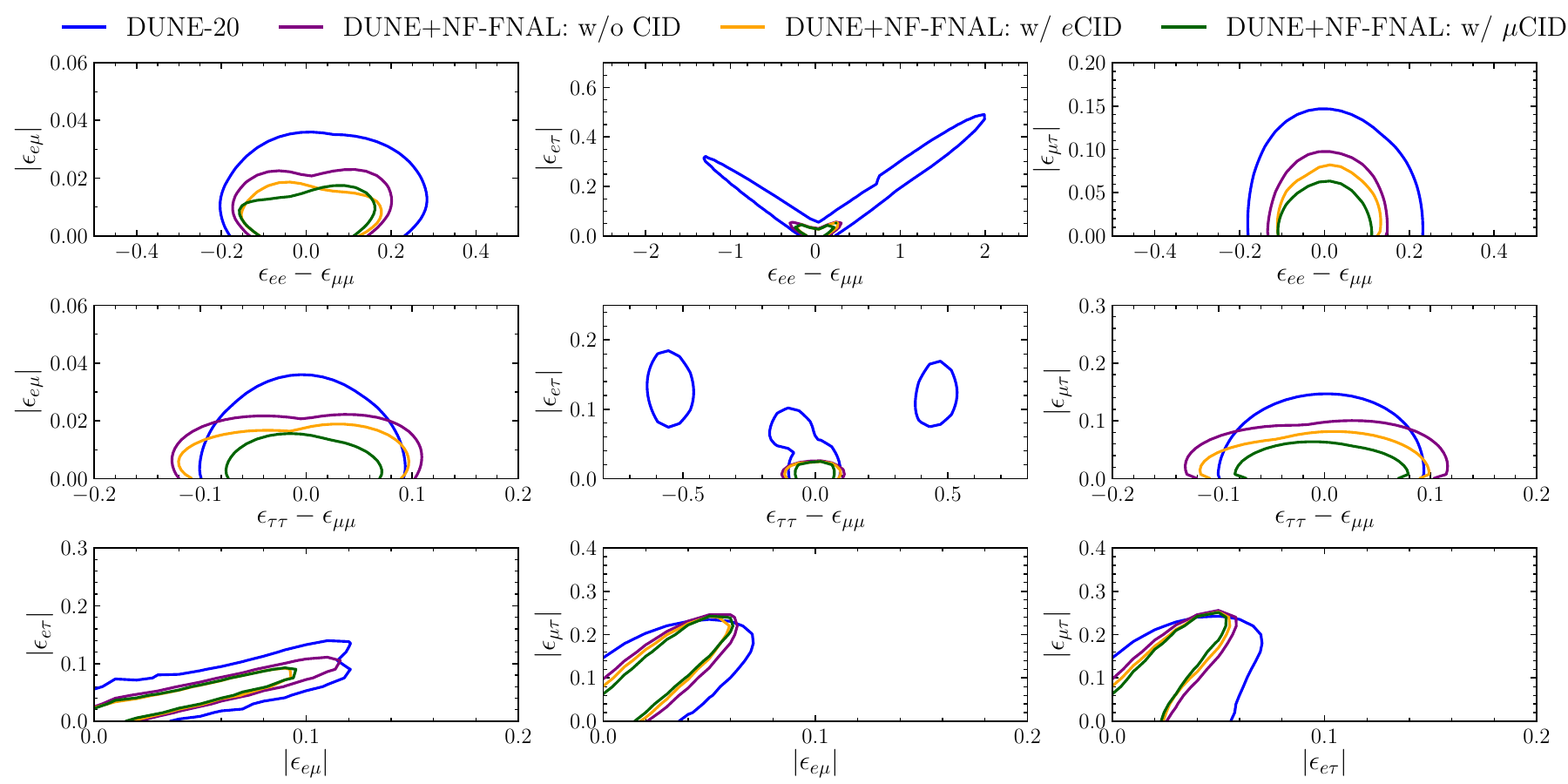}
\caption{The experimental constraint on 
different combinations of two NSI parameters
at 90\% confidence level, where the corresponding complex phase of NSI off-diagonal parameter and $\delta_{\rm CP}$ are marginalized over.
See Fig.~\ref{fig:NSI 2D phase BNL} in the appendix for the sensitivities for the BNL configuration.}
\label{fig:NSI 2D phase}
\end{figure}

\begin{figure}
\centering
\includegraphics[width=0.49\textwidth]{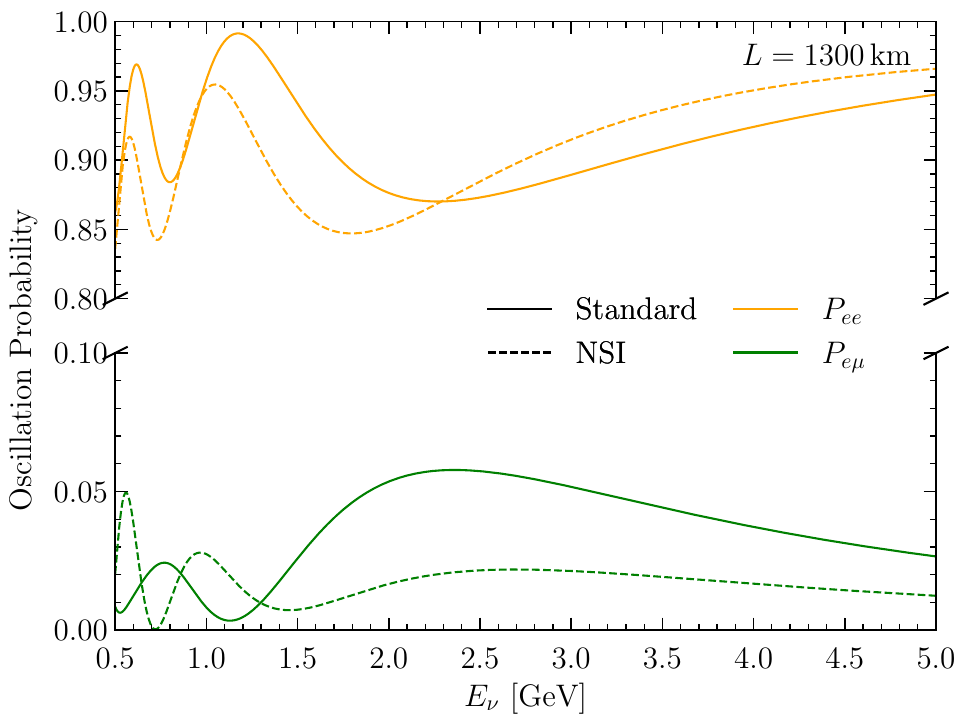}
\includegraphics[width=0.49\textwidth]{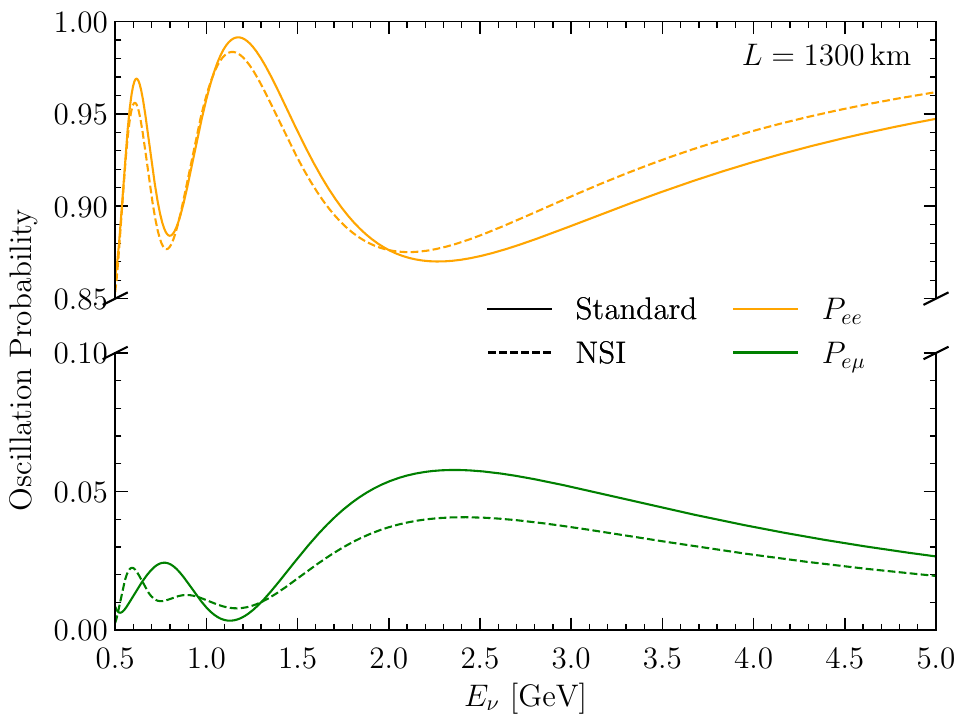}
\caption{The oscillation probability of $\nu_e\rightarrow\nu_e$ (orange), and $\nu_e\rightarrow\nu_\mu$ (green) with and without NSI using the best-fit values as summarized in Tab.~\ref{tab:oscillation-params}. In the NSI case, we take $\epsilon_{ee}-\epsilon_{\mu\mu} = 1.2, |\epsilon_{e\tau}| = 0.3, \phi_{e\tau} = -152^\circ$, $\theta_{23}=42^\circ$, and $\delta_{CP}=-53^\circ$ in the left panel, while we take $\epsilon_{\tau\tau}-\epsilon_{\mu\mu} = -0.55, |\epsilon_{e\tau}| = 0.125, \phi_{e\tau} = -133^\circ$, $\theta_{23}=45^\circ$, and $\delta_{CP}=-74^\circ$ in the right panel.
Note that in the NSI cases the values of oscillation parameters not specifically mentioned are the same as those listed in Tab.~\ref{tab:oscillation-params}.}
\label{fig:prob_epsee-epset}
\end{figure}

For a more general case, we take the off-diagonal NSI parameters to be complex and 
marginalize over the corresponding phases in Fig.~\ref{fig:NSI 2D phase}. 
Due to the free complex phase, the degeneracies can be
enlarged and significantly reduce the experimental 
sensitivities to the NSI parameters.
Notably, the degeneracy between $\epsilon_{ee}-\epsilon_{\mu\mu}$ and 
$|\epsilon_{e\tau}|$ leads to a wing-like structure 
for the original DUNE configuration which comes from the 
non-trivial dependence among the $\phi_{e\tau}$, $\delta_{CP}$ and $\theta_{23}$ parameters \cite{Coloma:2015kiu}. 
To better understand this degeneracy we show in the left panel of Fig.~\ref{fig:prob_epsee-epset} the oscillation probabilities without NSI and with non-zero $\epsilon_{ee}-\epsilon_{\mu\mu}$ and $\epsilon_{e\tau}$.
For the standard case without NSI, we use the values of oscillation parameters from Tab.~\ref{tab:oscillation-params}, while we take the best fit values that minimize the $\chi^2$ function for the NSI case.
More explicitly, we choose the point ($\epsilon_{ee}-\epsilon_{\mu\mu}=1.2$ and $|\epsilon_{e\tau}|=0.3$) in the wing-like structure panel of Fig.~\ref{fig:NSI 2D phase} as the input.
The $\chi^2$ minimum is achieved for the parameter $\phi_{e\tau} = -152^\circ$, $\theta_{23}=42^\circ$, and $\delta_{CP}=-53^\circ$, which indicates this parameter set has a degeneracy with the standard case without NSI. 
As DUNE only measures $\nu_e$ appearance and $\nu_\mu$ disappearance channels (and their anti-neutrino equivalents), this degeneracy significantly reduces DUNE's sensitivity to these NSI parameters.
The addition of a neutrino factory provides powerful new measurements to break this degeneracy and significantly enhance the sensitivity in the $\epsilon_{ee}-\epsilon_{\mu\mu}$-$|\epsilon_{e\tau}|$ parameter space beyond the naive expectations from statistics and new oscillation channels alone.
This is because a neutrino factory can measure the $\nu_e$ disappearance channel and the $\nu_\mu$ appearance channel, which help break the existing degeneracy in the $\nu_e$ appearance and $\nu_\mu$ disappearance channels.

Another important degeneracy is the island-like structure that occurs
for DUNE in the $\epsilon_{\tau\tau}-\epsilon_{\mu\mu}$ and $|\epsilon_{e\tau}|$ parameter space, see e.g.~\cite{Chatterjee:2021wac}.
This also arises from the non-trivial dependence on the  
$\phi_{e\tau}$, $\delta_{CP}$ and $\theta_{23}$ parameters as shown in the right panel of Fig.~\ref{fig:prob_epsee-epset}. 
Still, the $\nu_e$ disappearance channel and $\nu_\mu$ appearance channel can break the degeneracy in the original DUNE configuration. As a result, 
a neutrino factory can constrain these two parameters 
to a narrow region. For other combinations, notable 
improvements are also made by adding a neutrino factory.

In addition to the expected constraints on the NSI parameters,
we also study the precision of the Dirac CP phase as well as the sensitivity to
 $\theta_{23}$ in the presence of NSI.
To quantify the expected precision on to $\delta_{CP}$, we define $\Delta \delta_{CP}\equiv \rm max\left\{|\delta_{CP}^{\rm fit} - \delta_{CP}^{\rm true}|\right\}$ to be
the maximum absolute value of the difference between $\delta_{CP}^{\rm fit}$ and $\delta_{CP}^{\rm true}$,
where $\delta_{CP}^{\rm fit}$ is the fitted Dirac CP phase which gives $\Delta\chi^2 = 2.71$ (90\% C.L.) and $\delta_{CP}^{\rm true}$ is the fixed true Dirac CP phase.
We take the benchmark value of $\delta^{\rm true}_{CP}=0^\circ$ or $-90^\circ$, and three benchmark values of
$\theta^{\rm true}_{23}=42^\circ$, $45^\circ$, or $49^\circ$ for completeness. 
Similar to the standard oscillation framework without NSI, 
the precision on $\delta_{CP}^{\rm true}=0^\circ$ is better than for the $\delta_{CP}^{\rm true}=-90^\circ$ case while the octant of $\theta_{23}$ has only a mild impact on the CP precision. A sizable improvement exists for $\delta_{CP}^{\rm true}=-90^\circ$, as seen in Fig.~\ref{fig:Delta delta NSI 1D minimized}.
The results for $\delta^{\rm{true}}_{CP}=0$ are shown in the appendix which show qualitatively similar improvements to the precision of $\delta_{CP}$ with the addition of a neutrino factory.

\begin{figure}
\centering
\includegraphics[width=\textwidth]{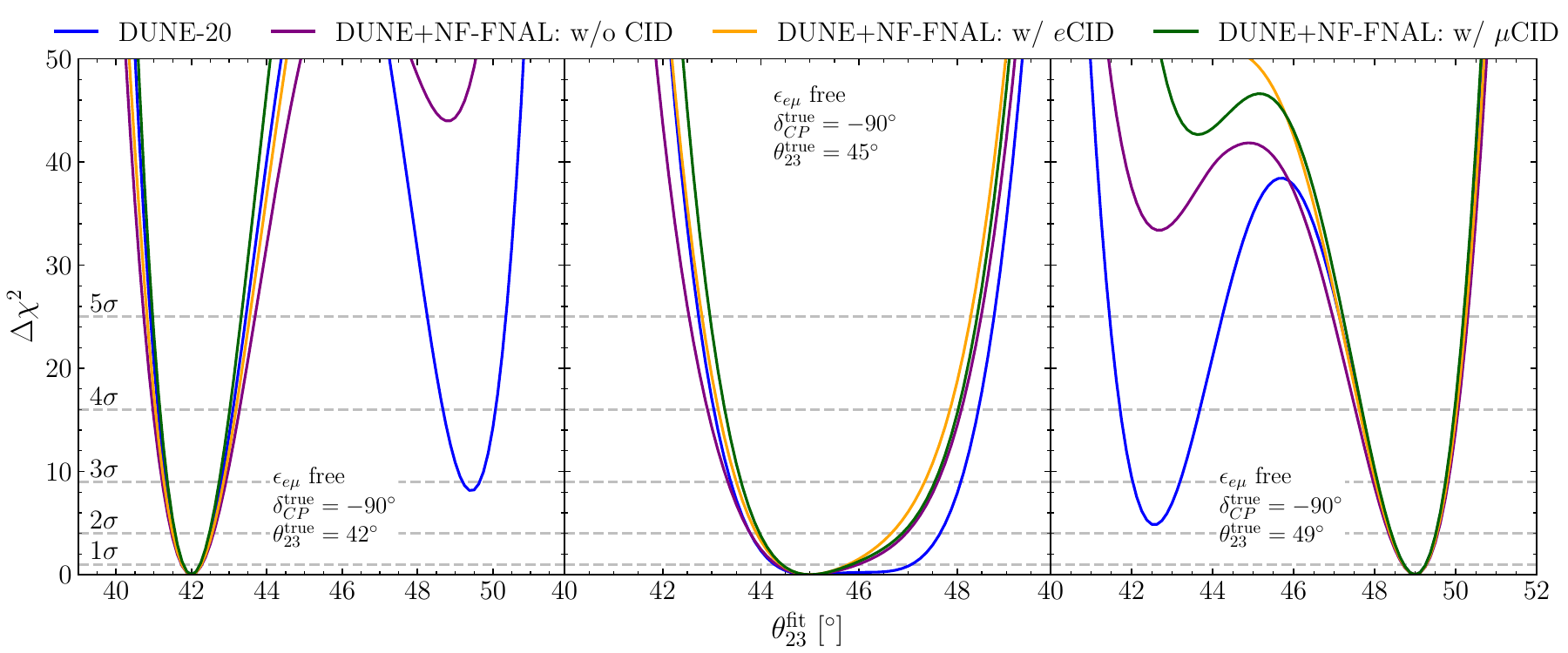}
\caption{The experimental sensitivity to $\theta_{23}$ for different true values of $\theta_{23}$ and we assume no NSI effect in the true case. For the fitting parameter, we consider the complex off-diagonal NSI parameter $\epsilon_{e\mu}$, which is minimized over including its complex phase.
See Figs.~\ref{fig:theta23 NSI em 1D BNL}-\ref{fig:theta23 NSI et 1D} in the appendix for the BNL configuration and other NSI parameter setups.}
\label{fig:theta23 NSI em 1D}
\end{figure}

In Fig.~\ref{fig:theta23 NSI em 1D} we show the sensitivity to $\theta_{23}$ for different benchmark 
values of $\delta^{\rm true}_{CP}$ and $\theta^{\rm true}_{23}$ as discussed above.
Here we take the $\epsilon_{e\mu}$ parameter for illustration since the neutrino factory has a significant improvement in this case,
while the results for other NSI parameters are shown 
in the appendix.
For this the figure, we assume $\epsilon_{e\mu}=0$.
We find that the sensitivity to $\theta_{23}$, for the benchmark values of $\theta_{23}^{\rm true}=42^\circ$ and $49^\circ$, at DUNE can be significantly reduced if one also considers NSI.
As a result, the octant of $\theta_{23}$ cannot be determined with over $\sim 3\sigma$ C.L.~at the benchmarks considered.
With the addition of a neutrino factory, however, the sensitivity to the octant of $\theta_{23}$ can reach over $5\sigma$.
We also find that the true value of $\delta_{CP}^\text{true}$ has only a small impact on the sensitivity.

\begin{figure}
\centering
\includegraphics[width=\textwidth]{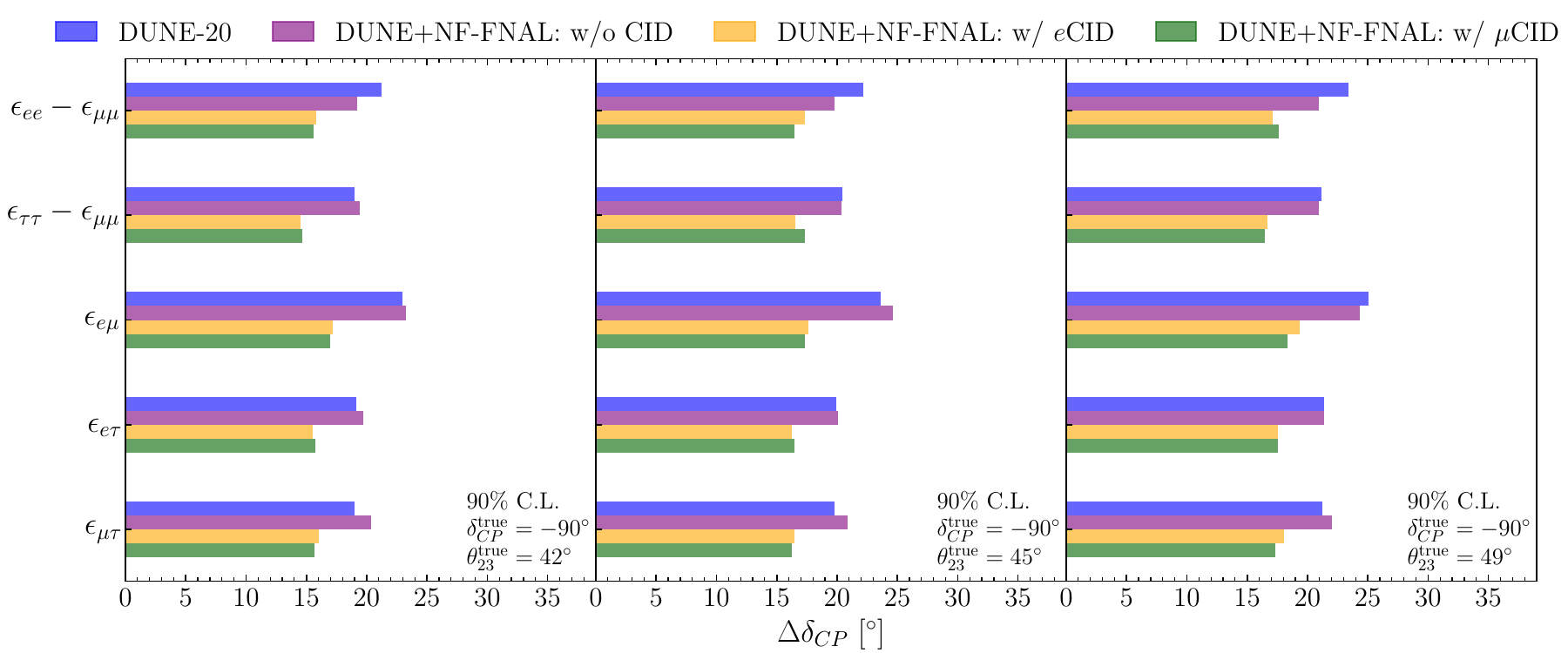}
\caption{The expected uncertainty on 
$\delta_{CP}$
at 90\% confidence level, where the corresponding NSI parameter is minimized over.
In this figure, the off-diagonal NSI parameters are taken as complex and the corresponding phase is minimized together with its modulus at a time. We assume $\delta_{CP}^\text{true}=-90^\circ$ and $\theta_{23}^\text{true}=42^\circ,~45^\circ,~49^\circ$.
See Fig.~\ref{fig:Delta delta NSI 1D minimized BNL} in the appendix for the BNL configuration and Fig.~\ref{fig:delta NSI 1D minimized} in the appendix for $\delta_{CP}=0$.}
\label{fig:Delta delta NSI 1D minimized}
\end{figure}

\section{CPT Violation}
\label{sec:cpt}
\subsection{Framework}

The CPT symmetry, which connects the three discrete symmetries: charge conjugation (C), parity (P),
and time reversal (T), is one of the most essential 
symmetries in particle physics and its invariance has been a guiding tool to build models. 
The CPT theorem is based on several basic assumptions on models,
including Lorentz invariance, hermiticity of the Hamiltonian, and local commutativity.
Despite its foundational nature, CPT can be violated in several classes of more complete models, some of which affect neutrino oscillations \cite{Kostelecky:1988zi,Kostelecky:1989jp,Kostelecky:1989jw,Sheikh-Jabbari:2000hdt,Carroll:2001ws,Greenberg:2002uu,Barenboim:2002tz,DeGouvea:2002xp,Ge:2019tdi}.
Certainly, the discovery of the CPT violation would 
indicate that at least one basic assumption should be
rejected and all our model building strategy need to be 
revisited.
Therefore, testing the CPT symmetry is one of the major tasks in particle physics. 

Neutrino oscillation provides a practical 
way to test CPT violation since a 
typical long-baseline neutrino experiment 
detects both neutrinos and anti-neutrinos.
As a consequence of CPT violation, different 
mass terms for neutrinos and anti-neutrinos can
be present, which allows for different mixing parameters. 
For clarity, we consider the flavor oscillations 
of neutrinos are described by the low-energy parameters $x\in \left\{\theta_{ij}, \delta_{CP}, \Delta m_{ij}^2\right\}$, while 
anti-neutrinos are associated with $\bar x \in \left\{\overline{\theta}_{ij}, \overline{\delta}_{CP}, \Delta \overline{m}_{ij}^2 \right\}$ where the $\theta_{ij}$ are the mixing angles, $\delta$ is the complex CP violating phase, and the $\Delta m^2$'s are the mass squared differences.
If CPT is conserved, the parameters should coincide 
except for a minus sign difference of the Dirac 
CP phase term.
The constraint on CPT violation has been analyzed using data taken from the accelerator (T2K and NO$\nu$A) and reactor (Daya Bay and RENO) neutrino experiments \cite{Tortola:2020ncu}.
A synergy between two on-going experiments (T2K and NO$\nu$A) and
one upcoming reactor experiment (JUNO) 
has been studied in \cite{Ngoc:2022uhg}.
A specific bound on the CPT violation in the solar neutrino sector has been studied using the KamLAND and solar neutrino data \cite{Barenboim:2023krl}. 
Moreover, projected constraints are discussed for next-generation experiments which include JUNO, DUNE, Hyper-Kamiokande, and ESSnuSB in \cite{Barenboim:2017ewj,deGouvea:2017yvn,Majhi:2021api,Barenboim:2023krl}. 

Since there are two sets of the neutrino oscillation parameters in the presence of
CPT violation,
we take the following $\chi^2$ function to calculate experimental sensitivities \cite{Barenboim:2023krl},
\begin{equation}
  \chi^2 
\equiv 
  \chi^2_\nu (x) + \chi^2_{\bar\nu} (\bar x).
\end{equation}
Here $x$ ($\bar x$) corresponds to the oscillation parameter set in the neutrino (anti-neutrino) oscillation probabilities.

\begin{figure}
\centering
\includegraphics[width=\textwidth]{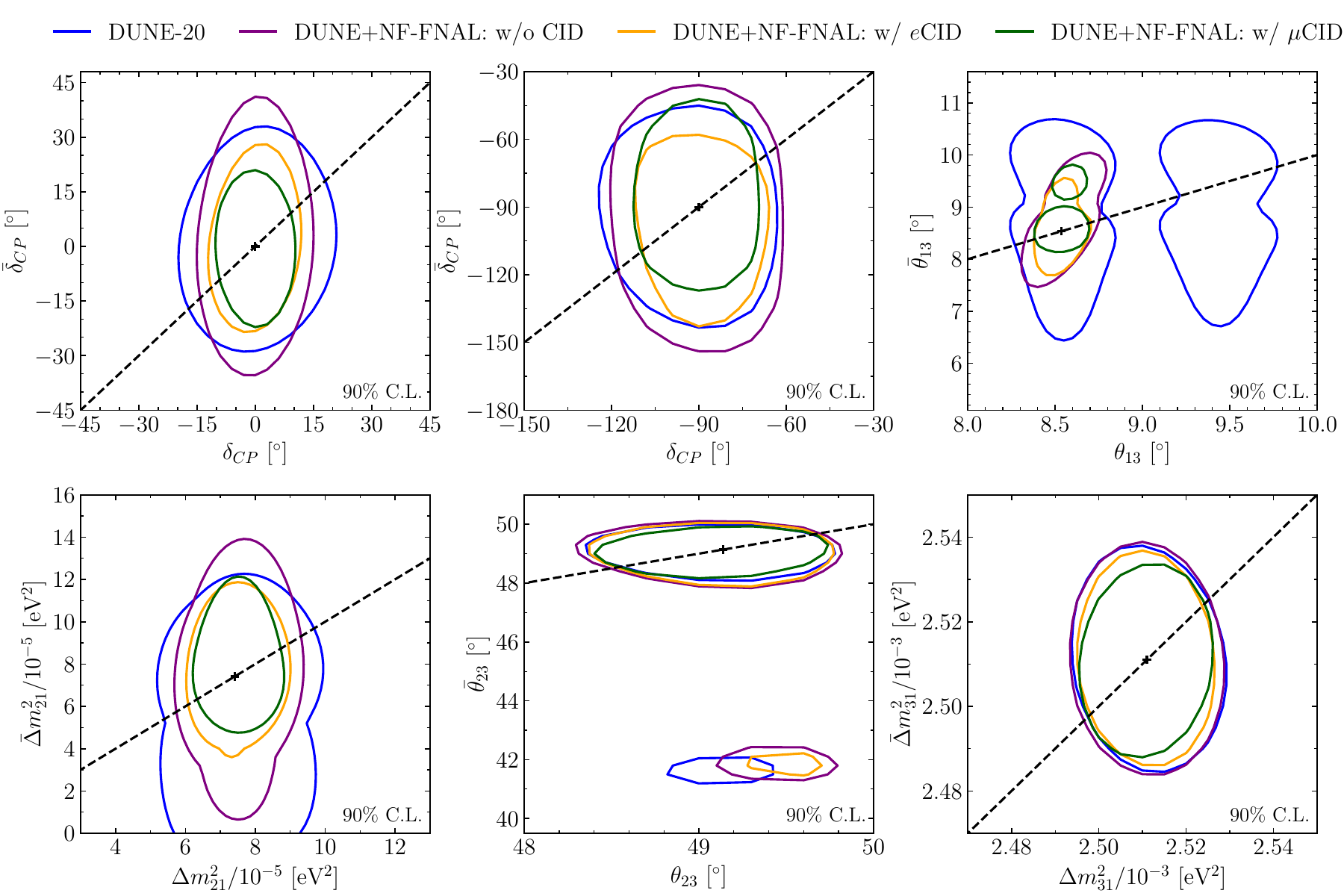}
\caption{The experimental sensitivity to the difference between the oscillation parameters for neutrino and anti neutrinos
$x$-$\bar x$
at 90\% confidence level for different experimental setups assuming CPT conservation. We mark the best-fit value with a star.
The dashed line shows the CPT conserving line. See Fig.~\ref{fig:CPT 2D BNL} in the appendix for the BNL configuration.}
\label{fig:CPT 2D}
\end{figure}

\begin{figure}
\centering
\includegraphics[width=0.49\textwidth]{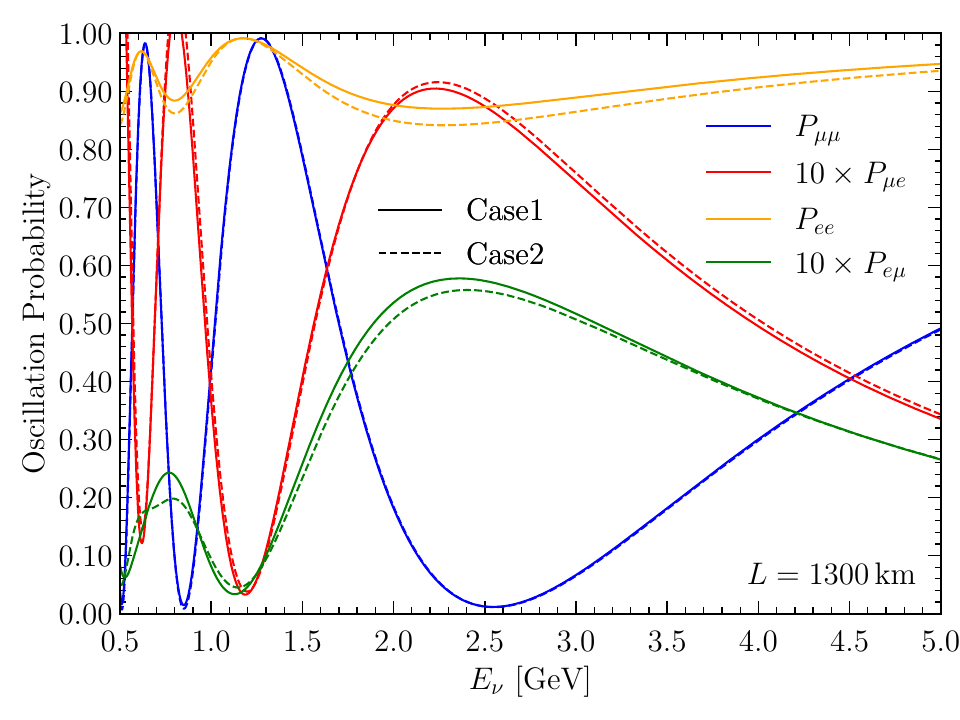}
\includegraphics[width=0.49\textwidth]{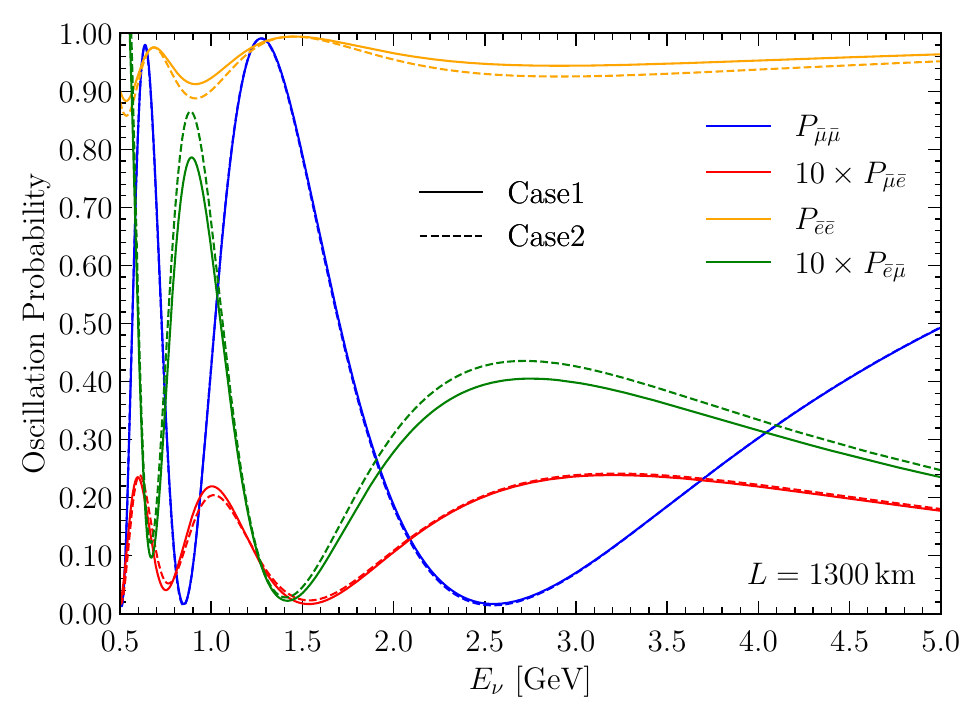}
\caption{\textbf{Left:} The oscillation probability of $\nu_\mu\rightarrow\nu_\mu$ (blue), $\nu_\mu\rightarrow\nu_e$ (red), $\nu_e\rightarrow\nu_e$ (orange), and $\nu_e\rightarrow\nu_\mu$ (green) for two oscillation parameter sets: case 1) the values of oscillation parameters are same as Tab.~\ref{tab:oscillation-params} (solid lines), and case 2) $\theta_{13}=9.5^\circ, \theta_{23}=43^\circ, \delta_{CP}=-83^\circ$ (dashed lines). 
\textbf{Right:} Same as the left panel but for the anti-neutrino mode and the parameter values in case 2: $\bar\theta_{23}=42^\circ, \bar\theta_{13}=9.9^\circ, \bar\delta_{CP}=-94^\circ$.
Note that the values of oscillation parameters not specifically mentioned are the same as those listed in Tab.~\ref{tab:oscillation-params}.}
\label{fig:prob_cpt}
\end{figure}

\begin{figure}
\centering
\includegraphics[width=0.95\textwidth]{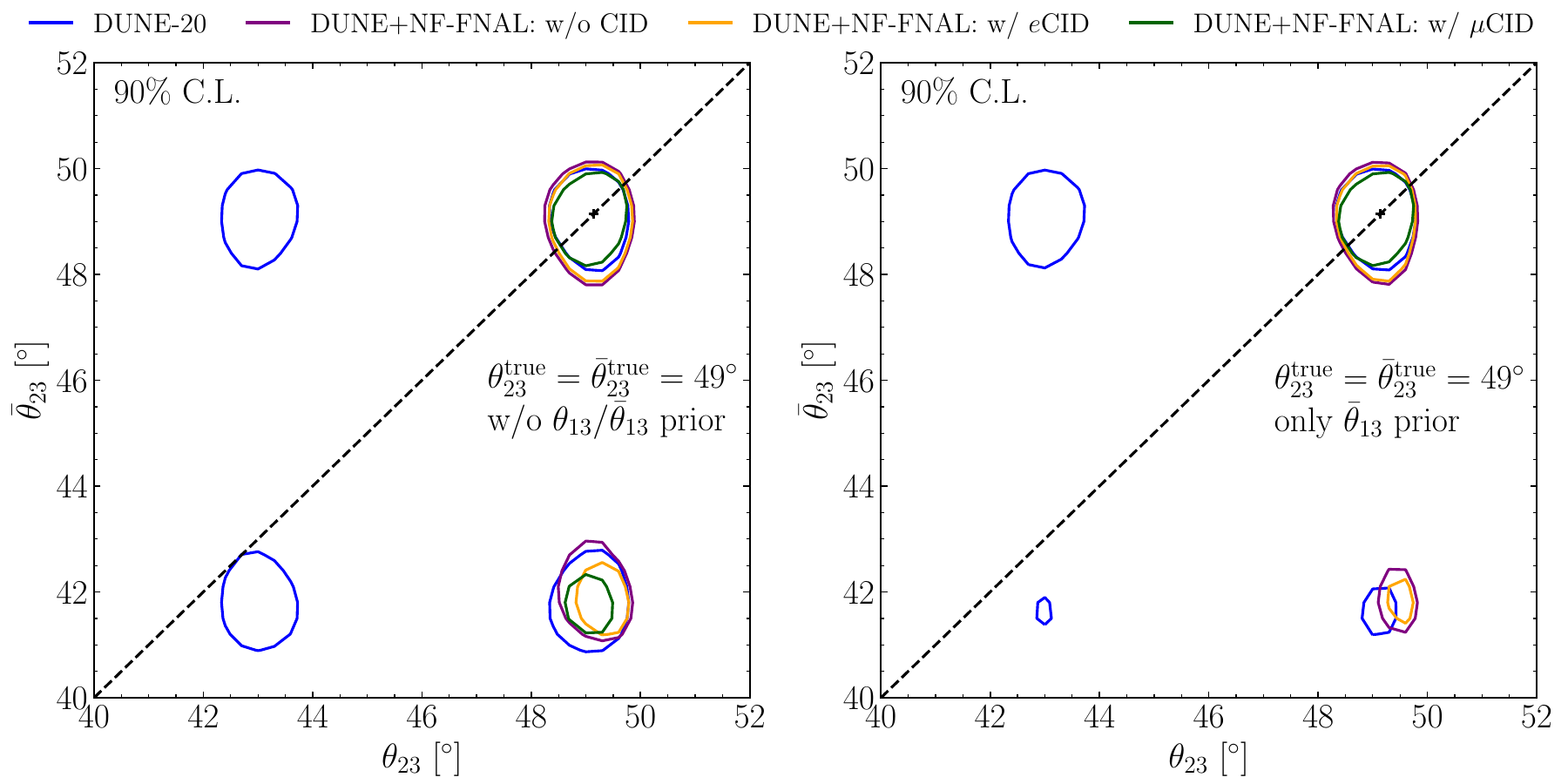}
\caption{The expected experimental constraint on 
$\theta_{23}$-$\bar \theta_{23}$
at 90\% confidence level. 
Left panel: both $\theta_{13}$ and 
$\bar\theta_{13}$ are free parameters
without any prior.
Right panel: both $\theta_{13}$ and 
$\bar\theta_{13}$ are free parameters, but only the prior of $\bar\theta_{13}$ is included. See Fig.~\ref{fig:CPT 2D zoom BNL} in the appendix for the BNL configuration.}
\label{fig:CPT 2D zoom}
\end{figure}

\begin{figure}
\centering
\includegraphics[width=0.95\textwidth]{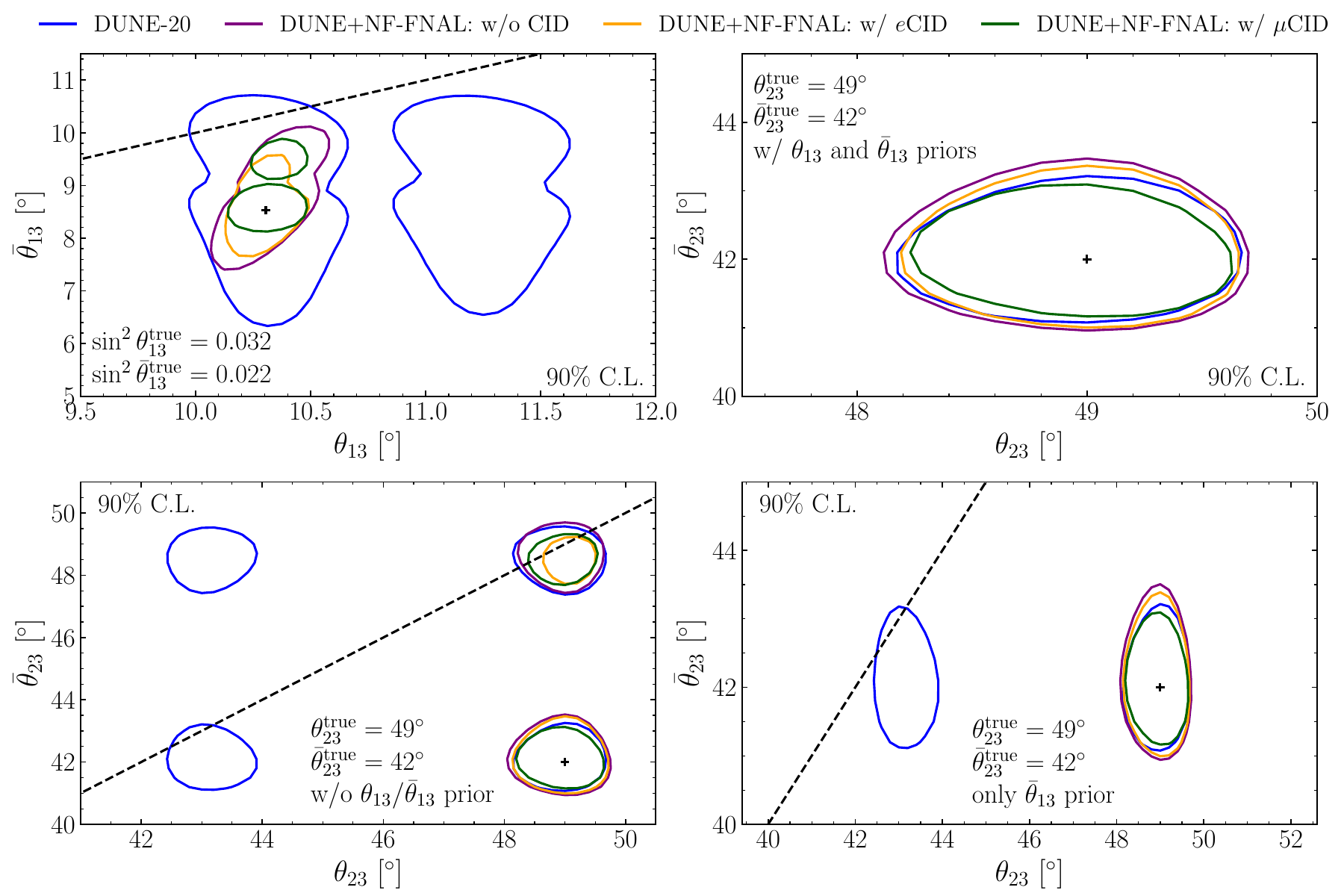}
\caption{\textbf{CPT violation benchmark plot.} Upper left panel: $\sin^2\theta_{13}^{\rm true} = 0.032, \sin^2\bar\theta_{13}^{\rm true} = 0.022$. 
Upper right panel: $\theta_{23}^{\rm true} = 49^\circ, \bar\theta_{23}^{\rm true} = 42^\circ$ with $\theta_{13}/\bar\theta_{13}$ priors.
Lower left panel: same as the upper right one but without $\theta_{13}/\bar\theta_{13}$ priors.
Lower right panel: same as the upper right one but only $\bar\theta_{13}$ prior is included.
See Fig.~\ref{fig:CPT 2D benchmark BNL} in the appendix for the BNL configuration.}
\label{fig:CPT 2D benchmark}
\end{figure}

\subsection{Results for the CPT framework}
We now present our numerical results on CPT violation sensitivities.
In Fig.~\ref{fig:CPT 2D}
we show the results for $x-\bar{x}$ at 90\% C.L.~for $x=\delta_{CP},~\theta_{13},~\theta_{23}$ and both mass splittings assuming $\delta_{CP}^\text{true}=0$ or $-90^\circ$ and no CPT violation. Since a long-baseline experiment is not very sensitive to the solar mixing angle $\theta_{12}$, we do not show the CPT results for it for simplicity.
In general, the addition of a NF to DUNE improves the precision on the oscillation parameters for neutrinos and anti-neutrinos simultaneously, in particular when CID is implemented. An exception to this is the results for $\Delta m_{31}^2$ which are similar for 20 years of DUNE and DUNE+NF, nearly independent of the choice of CID.
We find that the neutrino oscillation parameters can be more precisely constrained than the anti-neutrino parameters due to the larger statistics.

Accelerator experiments do not have a large sensitivity to the solar parameters; they are most precisely determined by solar and reactor experiments. Concerning CPT violation we see that DUNE alone cannot exclude non-zero $\bar{\Delta} m_{21}^2$ at 90\% C.L.~while this parameter in neutrino mode can be constrained to be non-zero. Adding a neutrino factory, in particular with the FNAL baseline allows to exclude zero $\bar{\Delta}m_{21}^2$.

For $\theta_{13}$ DUNE only cannot distinguish $\theta_{13}=8.5^\circ$ from the $\theta_{13}=9.5^\circ$ degenerate point at 90\% C.L.~under the assumption of CPT violation even when doubling the running time. This is because the free $\theta_{23}$ angle and CP phase can degenerate with $\theta_{13}$. As shown in the left panel of Fig.~\ref{fig:prob_cpt}, for $\nu_\mu\rightarrow\nu_e$ and $\nu_\mu\rightarrow\nu_\mu$ channels which are 
main signals at DUNE, the oscillation probabilities with $(\theta_{13}=9.5^\circ$, $\theta_{23}=43^\circ$, $\delta_{CP}=-83^\circ)$ are very close to that of $(\theta_{13}=8.5^\circ, \theta_{23}=49^\circ, \delta_{CP}=-90^\circ)$, leading to the island-like structure in Fig.~\ref{fig:CPT 2D}. However, the $\nu_e\rightarrow\nu_e$ channel is almost independent of the CP phase and $\theta_{23}$ angle, which means a measurement of $\nu_e$ disappearance channel can significantly improve the $\theta_{13}$ sensitivity, in a similar fashion to a typical medium baseline reactor neutrino experiment, except with the added channel of $\nu_e\to\nu_e$ instead of just $\bar\nu_e\to\bar\nu_e$. The precision on $\bar{\theta}_{13}$ is $\pm 2.5^\circ$. 
The addition of a NF excludes the $\theta_{13}=9.5^\circ$ degeneracy and the precision on $\bar{\theta}_{13}$ also improves.

For $\bar{\theta}_{23}$ DUNE alone or combined with a NF cannot resolve the octant at 90\% C.L.~in the CPT violating scenario unless the NF has good $\mu$CID. 
As shown in the right panel of Fig.~\ref{fig:prob_cpt}, the $\nu_\mu$ disappearance channel is sensitive to $\sin^22\theta_{23}$ and hence not sensitive to the $\theta_{23}$ octant.
However, appearance channels are sensitive to $\sin^2\theta_{23}$, so the octant can be resolved.
In the case of a neutrino factory, $\mu$CID is required to provide a clean measurement of the $\nu_\mu$ appearance channel.
Consequently, the sensitivity of $\theta_{23}$ octant can be largely improved in the presence of CPT violation.
The octant can be resolved for the neutrino parameter due to the larger statistics in the neutrino channel.

Of interest is also the impact of $\theta_{13}$ in the determination of $\theta_{23}$ octant because the main contribution to the oscillation probability $P_{\mu e}$ is $4\sin^2\theta_{23}\sin^2\theta_{13}\cos^2\theta_{13}\sin^2\Delta_{31}$ with $\Delta_{31}\equiv \Delta m^2_{31}L/4E_\nu$, which means $\theta_{13}$ can affect the $\theta_{23}$ octant. 
In Fig.~\ref{fig:CPT 2D zoom}
we study the expected constraints without using an external prior on $\theta_{13}$ or $\bar{\theta}_{13}$. DUNE alone cannot resolve the octant for either $\theta_{23}$ or $\bar{\theta}_{23}$. When a NF is added only two degenerate solutions for $\bar{\theta}_{23}$ survive due to the lower statistics in the anti-neutrino channel. The prior on $\bar{\theta}_{13}$ from reactor anti-neutrino measurements is the driving force behind removing the wrong octant solutions for $\theta_{23}$. Having a prior on $\bar{\theta}_{13}$ also allows to remove the wrong-octant solution for $\bar{\theta}_{23}$ for a NF with $\mu$CID as discussed above. 

So far there is no experimental evidence that CPT is violated in nature. However, there is a mild discrepancy in the measurement
of $\theta_{13}$ and $\bar\theta_{13}$ coming from the measurement of solar \cite{Super-Kamiokande:2023jbt} and accelerator \cite{T2K:2023smv} neutrinos compared to reactor anti-neutrinos \cite{DayaBay:2022orm,Shin:2020mue,Soldin:2024fgt}.
We now consider benchmark values of $\sin^2\theta_{13}=0.032,~\sin^2\bar{\theta}_{13}=0.022$, which are consistent with the solar neutrino measurement and reactor neutrino measurement, respectively.
To study the impact of CPT for $\theta_{13}$ we show in Fig.~\ref{fig:CPT 2D benchmark} the expected sensitivities to $\theta_{13}$ and $\bar{\theta}_{13}$ assuming the true values to be CPT violating.
We again see that the addition of a NF excludes the higher solution for $\theta_{13}$ and the precision on both parameters is drastically improved due to the measurement of $\nu_e$ disappearance channel as discussed above; this would allow for the discovery of CPT violation if this hint was realized in nature.
We now consider a similar scenario but with the octant of $\theta_{23}$.
If we assume that $\theta_{23}^\text{true}=49^\circ,~ \bar{\theta}_{23}^\text{true}=42^\circ$ we again find that having a prior on $\bar{\theta}_{13}$ is crucial to exclude the wrong-octant solutions with a NF.
In some cases a CPT violating signal in the octant could be discovered with a NF given the inclusion of other oscillation data.

Finally, we comment on existing constraints from oscillations.
Long-baseline experiments will not be competitive with the measurements of the solar parameters $\theta_{12}$ and $\Delta m^2_{21}$ relative to the combined analysis of solar and reactor data which have more precision in the standard case \cite{Denton:2023zwa} and have the benefit of one class of experiments measuring neutrinos (solar) and the other measuring antineutrinos (reactors) \cite{Barenboim:2023krl}.
DUNE and a NF will improve on the constraints for the other four parameters considerably as shown in fig.~\ref{fig:CPT 2D}.
There are no constraints now on CPT violation appearing in $\delta$ as $\delta$ has not yet been measured in the CPT conserving case, so the sensitivity of LBL experiments is world-leading.
As for the other three parameters, the existing constraints at 90\% are at $|\Delta\sin^2\theta_{23}|\lesssim0.13$, $|\Delta\sin^2\theta_{13}|\lesssim0.019$, and $|\Delta(\Delta m^2_{31})|\lesssim0.14\e{-3}$ eV$^2$ \cite{Tortola:2020ncu} where $\Delta x\equiv x-\bar x$.
In comparison, we find that long-baseline accelerator experiments can achieve $|\Delta\sin^2\theta_{23}|\lesssim0.02$, $|\Delta\sin^2\theta_{13}|\lesssim0.007$, and $|\Delta(\Delta m^2_{31})|\lesssim0.03\e{-3}$ eV$^2$ where we have assumed that the relevant degeneracies can be lifted via the inclusion of data for $\bar\theta_{13}$.

\section{Conclusion}
\label{sec:conclusions}
We have studied the expected sensitivity of a neutrino factory to two new physics scenarios, vector new neutrino interactions and CPT violation.
A neutrino factory with a long baseline and LArTPC far detectors is particularly well suited to study these new physics scenarios due to its long baseline, excellent 
energy reconstruction (including potential charge identification abilities), and access to six oscillation channels plus their CP conjugate channels.

We find that combining 10 years of DUNE with 10 years of a neutrino factory will improve over the constraints on these new physics scenarios from DUNE alone even when considering 20 years of DUNE. Furthermore, the addition of a neutrino factory can remove numerous degeneracies in the DUNE-only constraints due to the presence of new oscillation channels, especially the $\nu_e$ disappearance and $\nu_\mu$ appearance channels.

\acknowledgments
PBD acknowledges support by the United States Department of Energy under Grant Contract No.~DE-SC0012704. JG acknowledges support by the U.S. Department of Energy Office of Science under
award number DE-SC0025448.
CFK acknowledges support by the National Natural Science Foundation of China (12425506, 12375101, 12090060
and 12090064) and the SJTU Double First Class 
start-up fund (WF220442604).
CFK also acknowledges the usage of INPAC cluster at Shanghai Jiao Tong University.

\appendix
\section{Additional NSI results}
\begin{figure}
\centering
\includegraphics[height=0.37\textwidth]{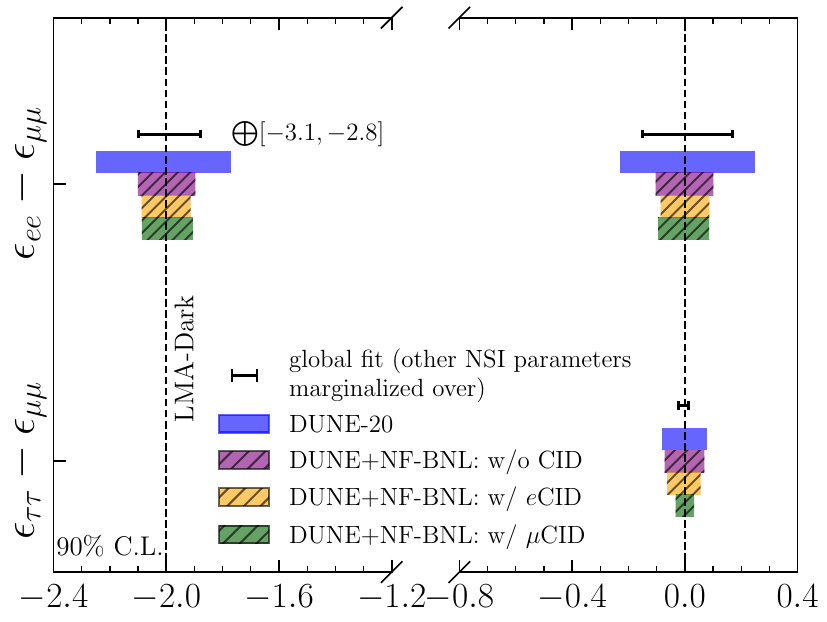}
\includegraphics[height=0.37\textwidth]{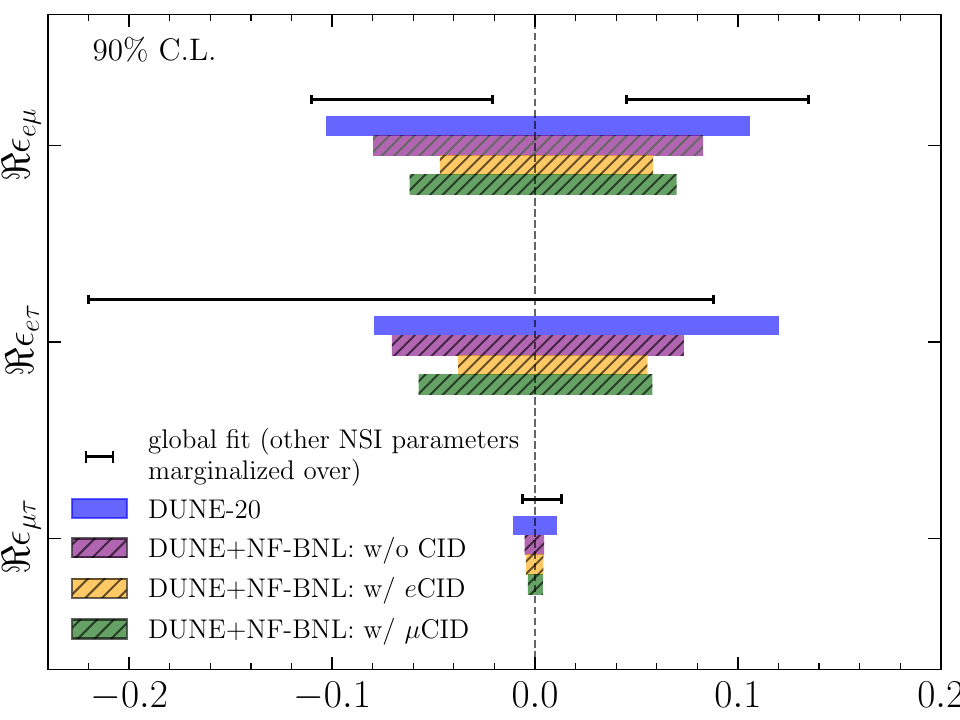}
\caption{Same as Fig.~\ref{fig:NSI 1D} but for the BNL configuration.}
\label{fig:NSI 1D BNL}
\end{figure}

\begin{figure}
\centering
\includegraphics[width=1\textwidth]{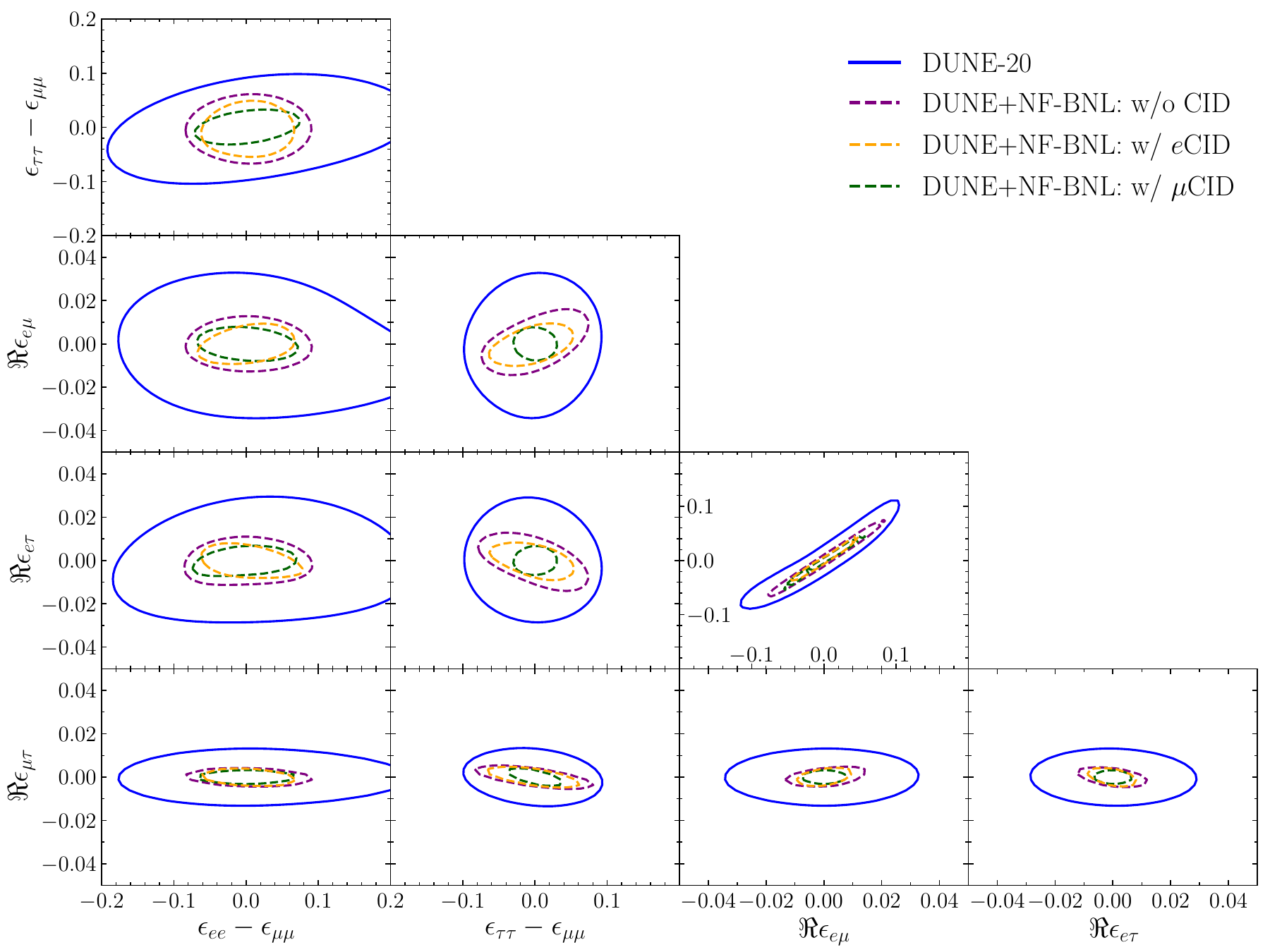}
\caption{Same as Fig.~\ref{fig:NSI two-D} but for the BNL configuration.}
\label{fig:NSI two-D BNL}
\end{figure}

\begin{figure}
\centering
\includegraphics[width=\textwidth]{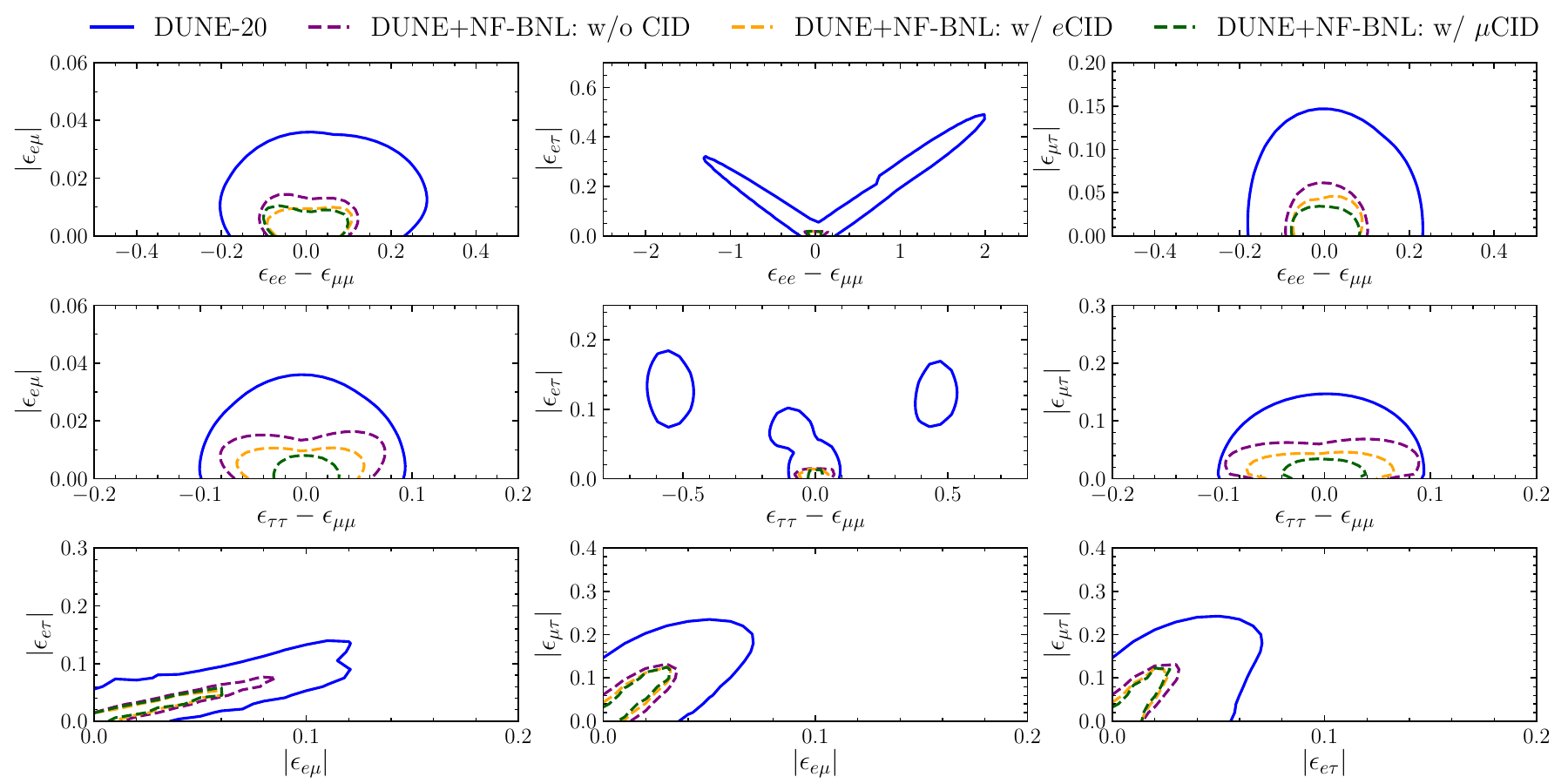}
\caption{Same as Fig.~\ref{fig:NSI 2D phase} but for the BNL configuration.}
\label{fig:NSI 2D phase BNL}
\end{figure}

\begin{figure}
\centering
\includegraphics[width=\textwidth]{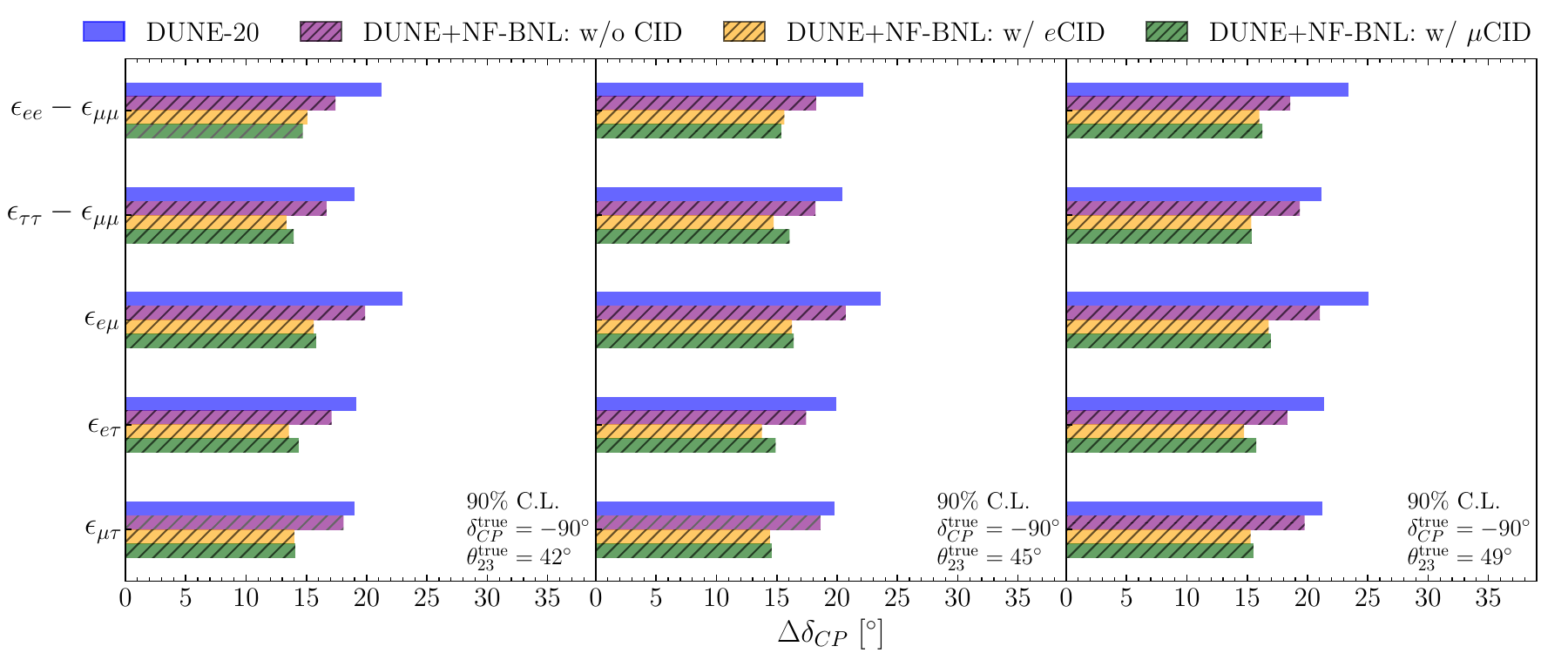}
\caption{Same as Fig.~\ref{fig:Delta delta NSI 1D minimized} but for the BNL configuration.}
\label{fig:Delta delta NSI 1D minimized BNL}
\end{figure}

\begin{figure}
\centering
\includegraphics[width=\textwidth]{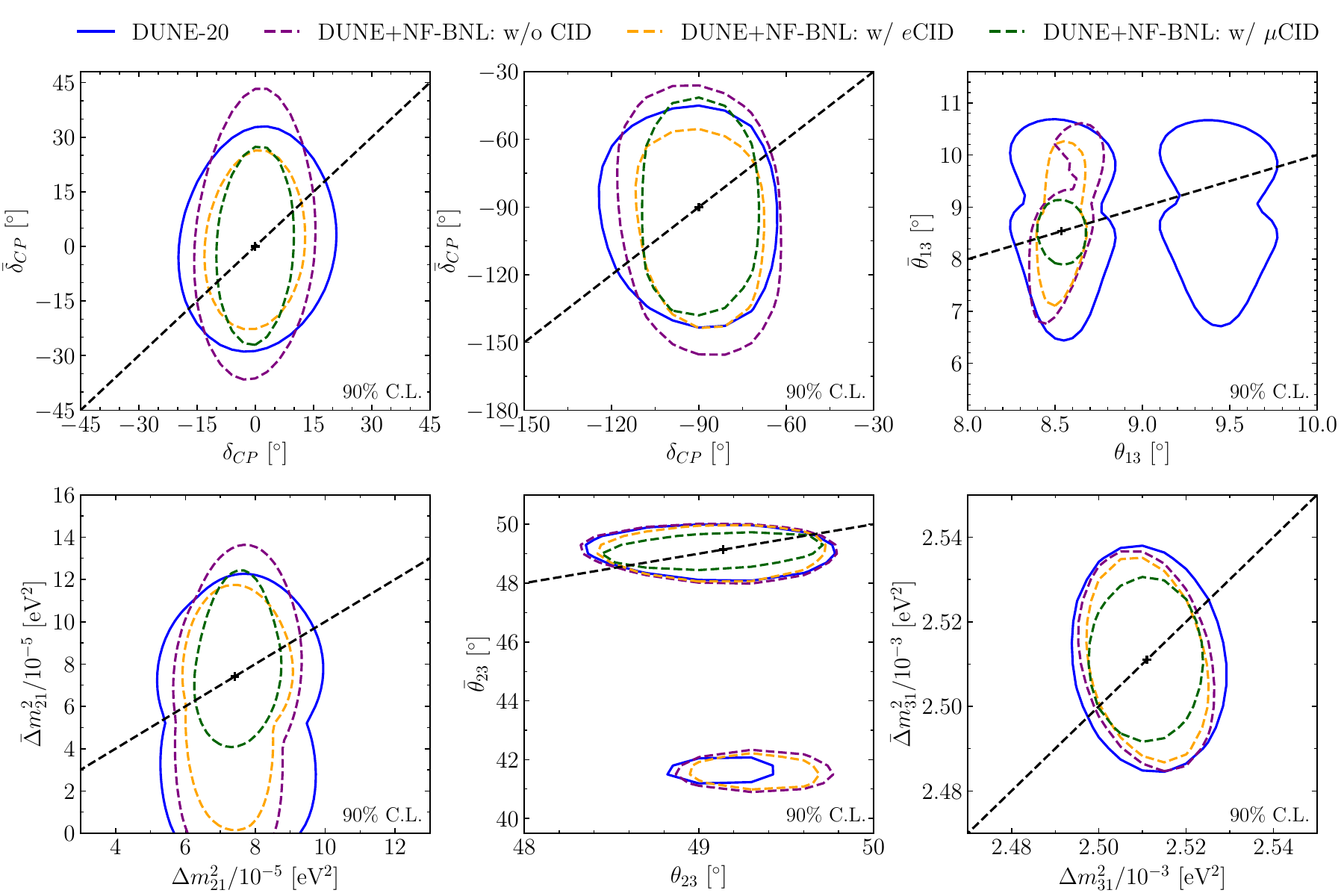}
\caption{Same as Fig.~\ref{fig:CPT 2D} but for the BNL configuration.}
\label{fig:CPT 2D BNL}
\end{figure}

\begin{figure}
\centering
\includegraphics[width=0.95\textwidth]{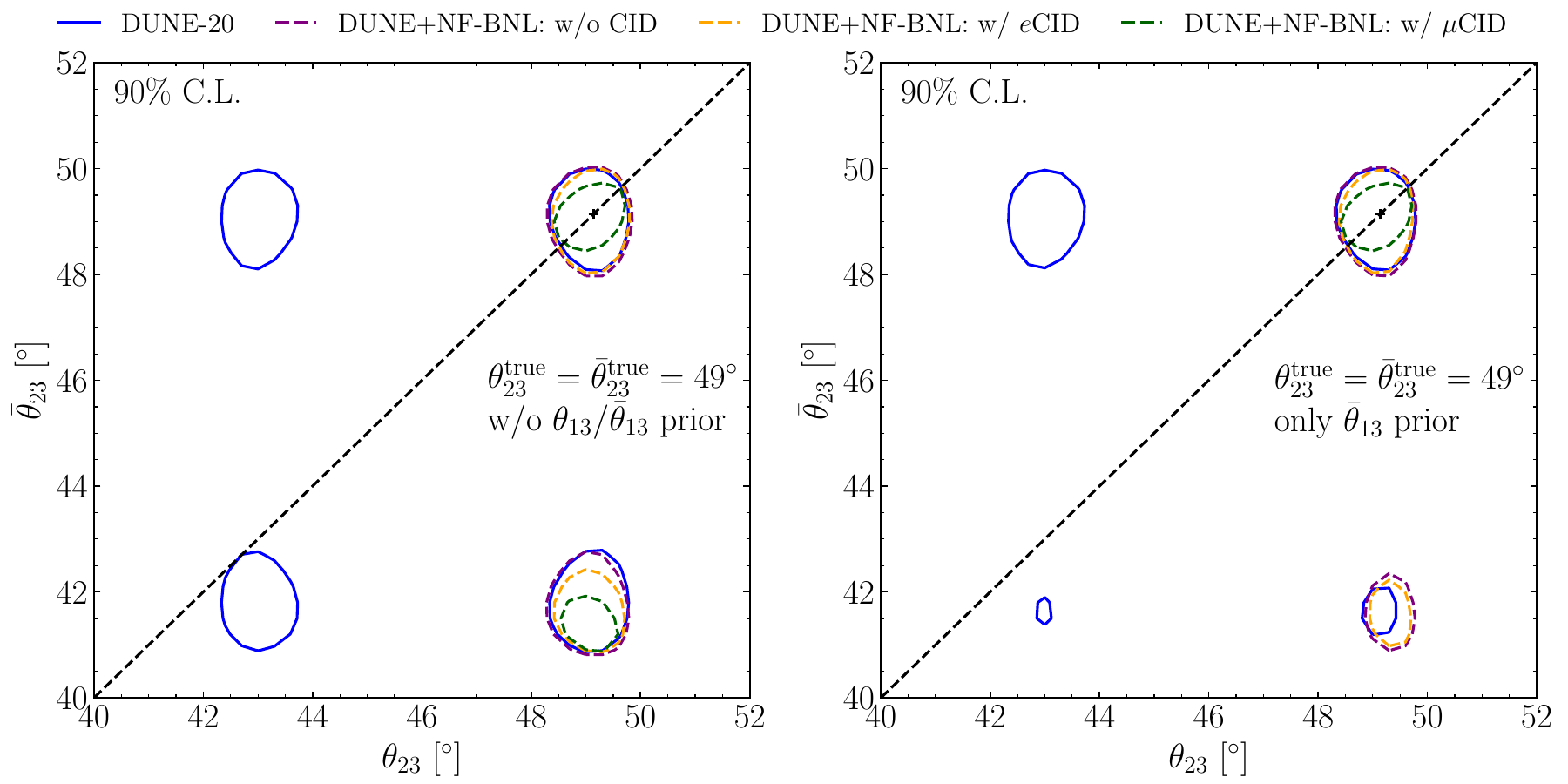}
\caption{Same as Fig.~\ref{fig:CPT 2D zoom} but for the BNL configuration.}
\label{fig:CPT 2D zoom BNL}
\end{figure}

\begin{figure}
\centering
\includegraphics[width=0.95\textwidth]{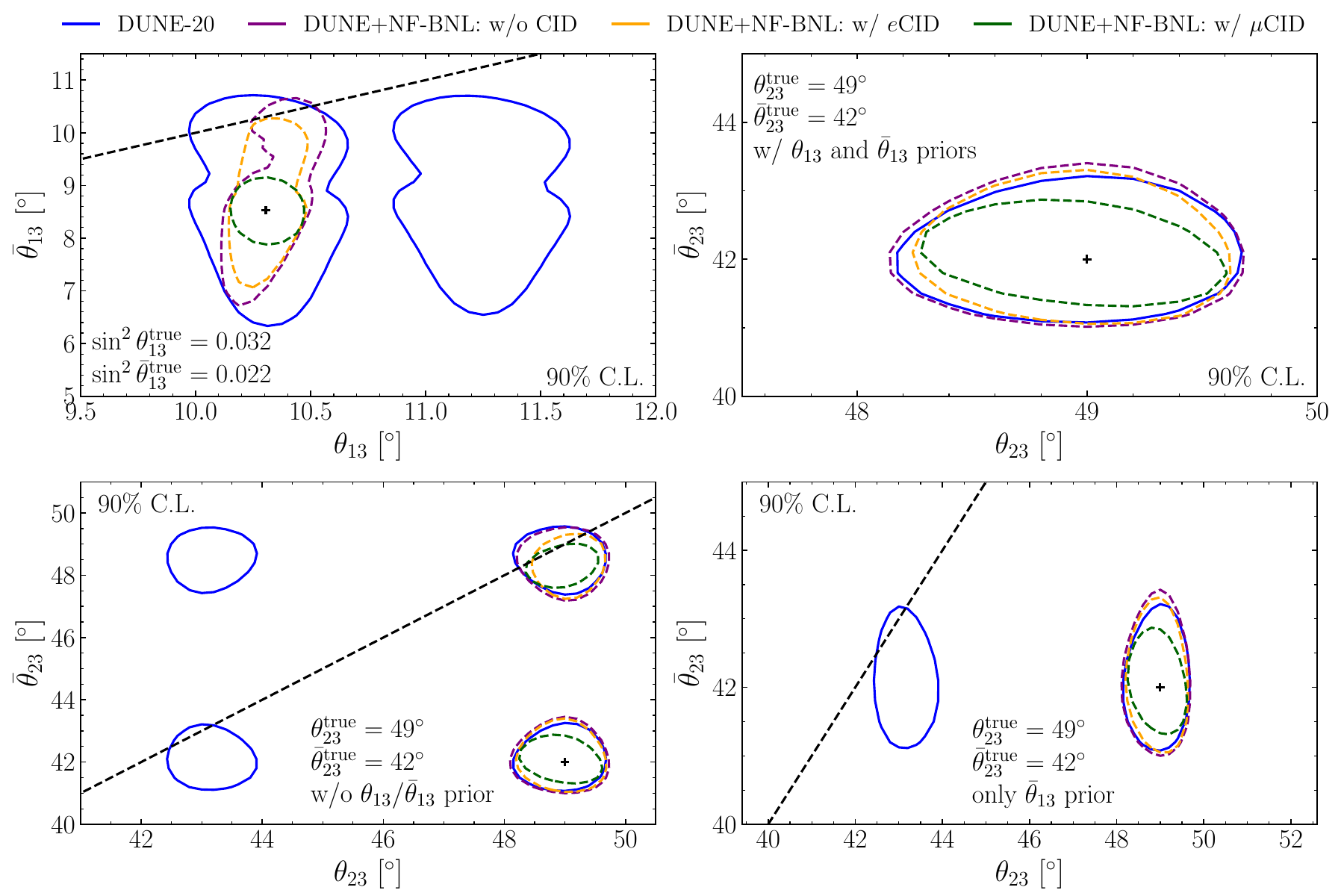}
\caption{Same as Fig.~\ref{fig:CPT 2D benchmark} but for the BNL configuration.}
\label{fig:CPT 2D benchmark BNL}
\end{figure}

\begin{figure}
\centering
\includegraphics[width=\textwidth]{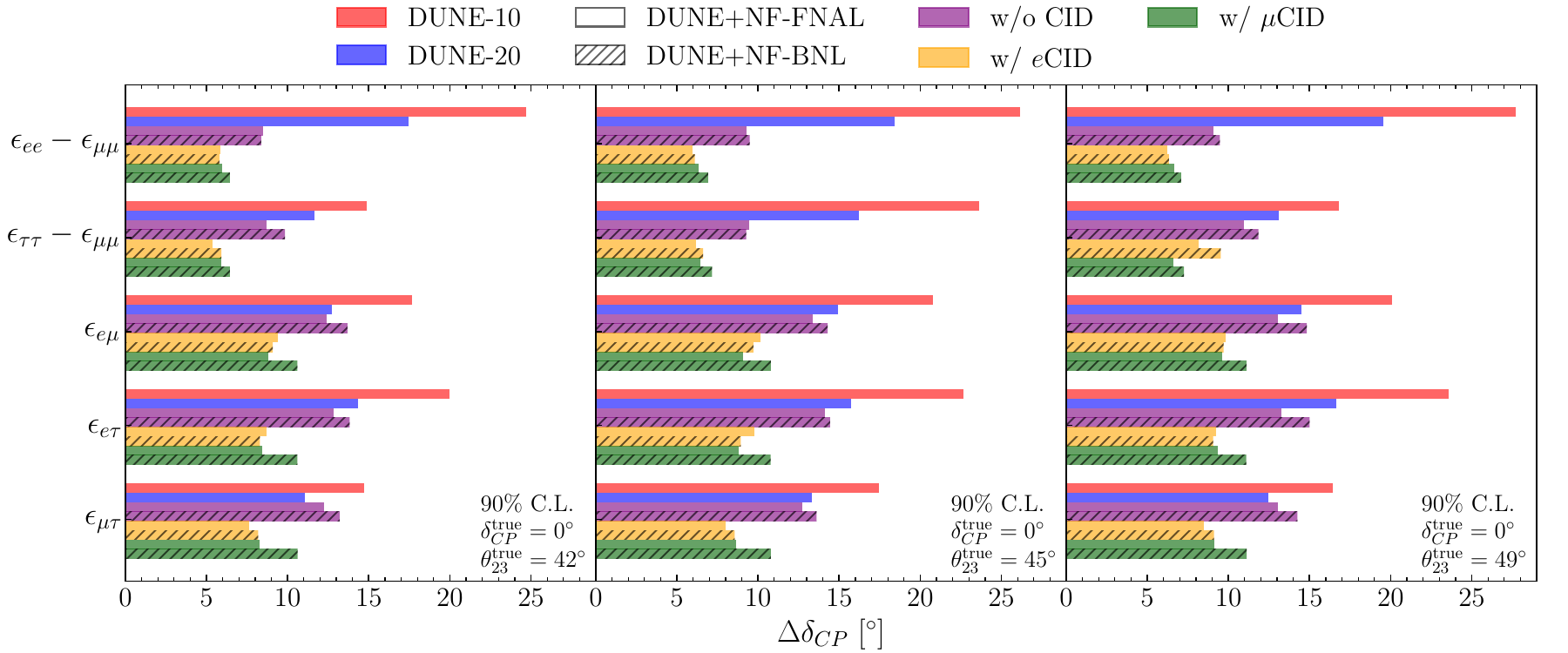}
\caption{The expected uncertainty on 
$\delta_{CP}$ assuming $\delta_{CP}^{\rm{true}}=0$
at 90\% confidence level, where the corresponding NSI parameter is minimized at a time. In this figure, the off-diagonal NSI parameters 
are taken as complex and the corresponding phase is minimized together with its modulus at a time. See Fig.~\ref{fig:Delta delta NSI 1D minimized} in the main text for $\delta_{CP}=-90^\circ$.}
\label{fig:delta NSI 1D minimized}
\end{figure}

\begin{figure}
\centering
\includegraphics[width=\textwidth]{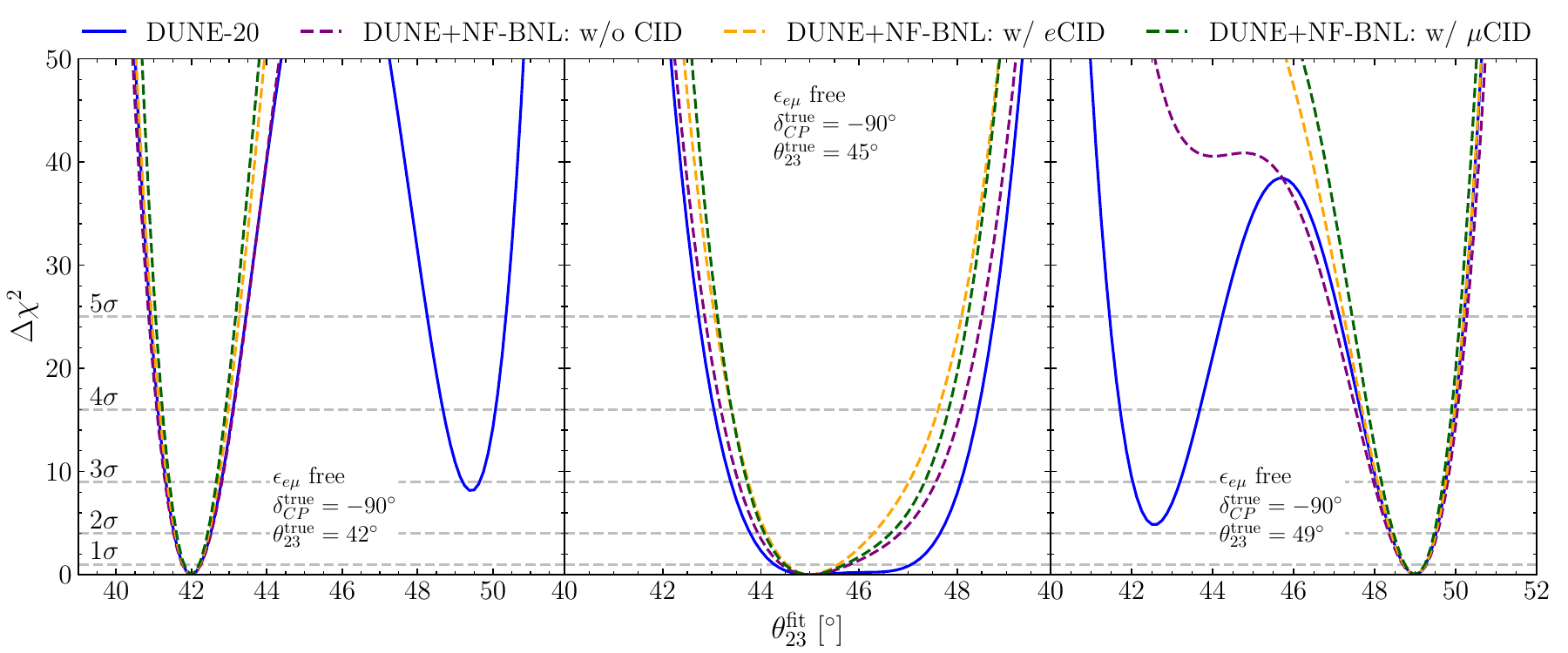}
\caption{Same as Fig.~\ref{fig:theta23 NSI em 1D} but for the BNL configuration.}
\label{fig:theta23 NSI em 1D BNL}
\end{figure}

\begin{figure}
\centering
\includegraphics[width=\textwidth]{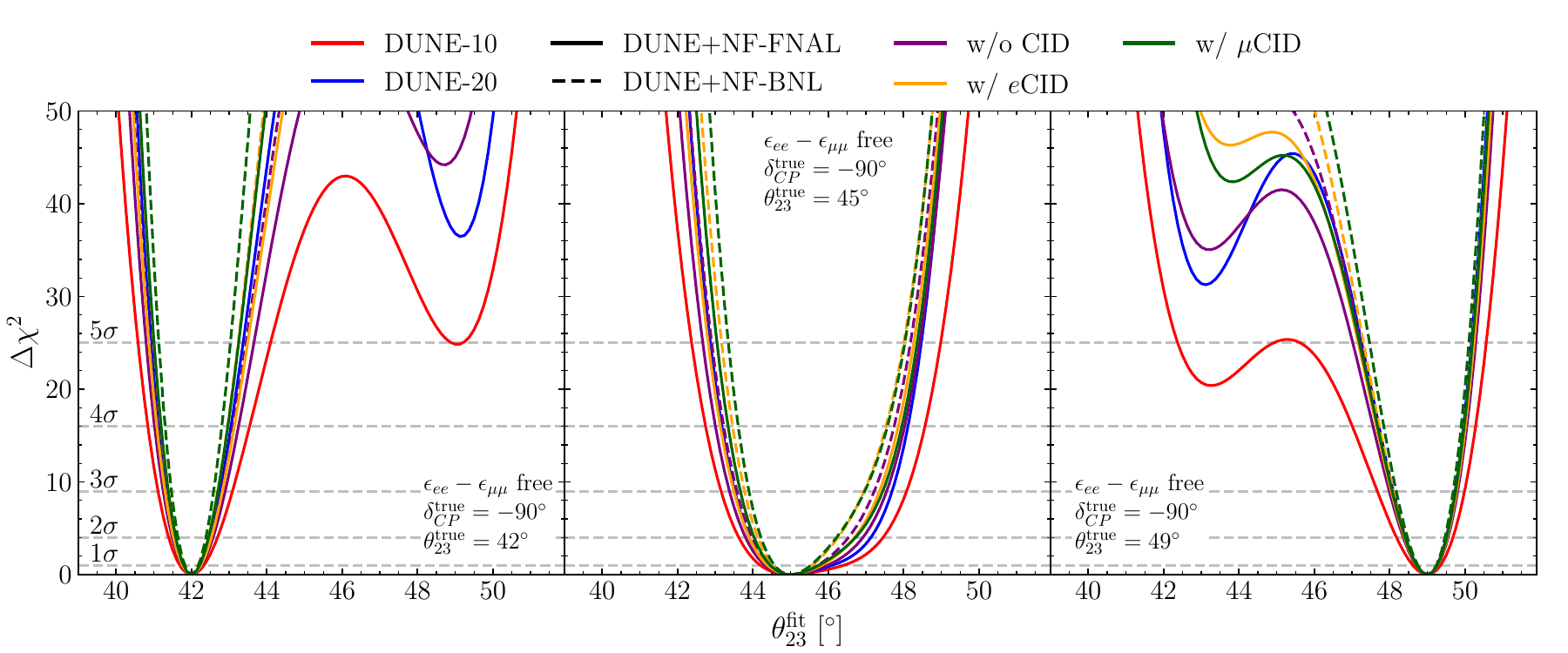}
\caption{The experimental sensitivity to $\theta_{23}$, where the $\epsilon_{ee}-\epsilon_{\mu\mu}$ is minimized at a time.}
\label{fig:theta23 NSI ee-mm 1D}
\end{figure}

\begin{figure}
\centering
\includegraphics[width=\textwidth]{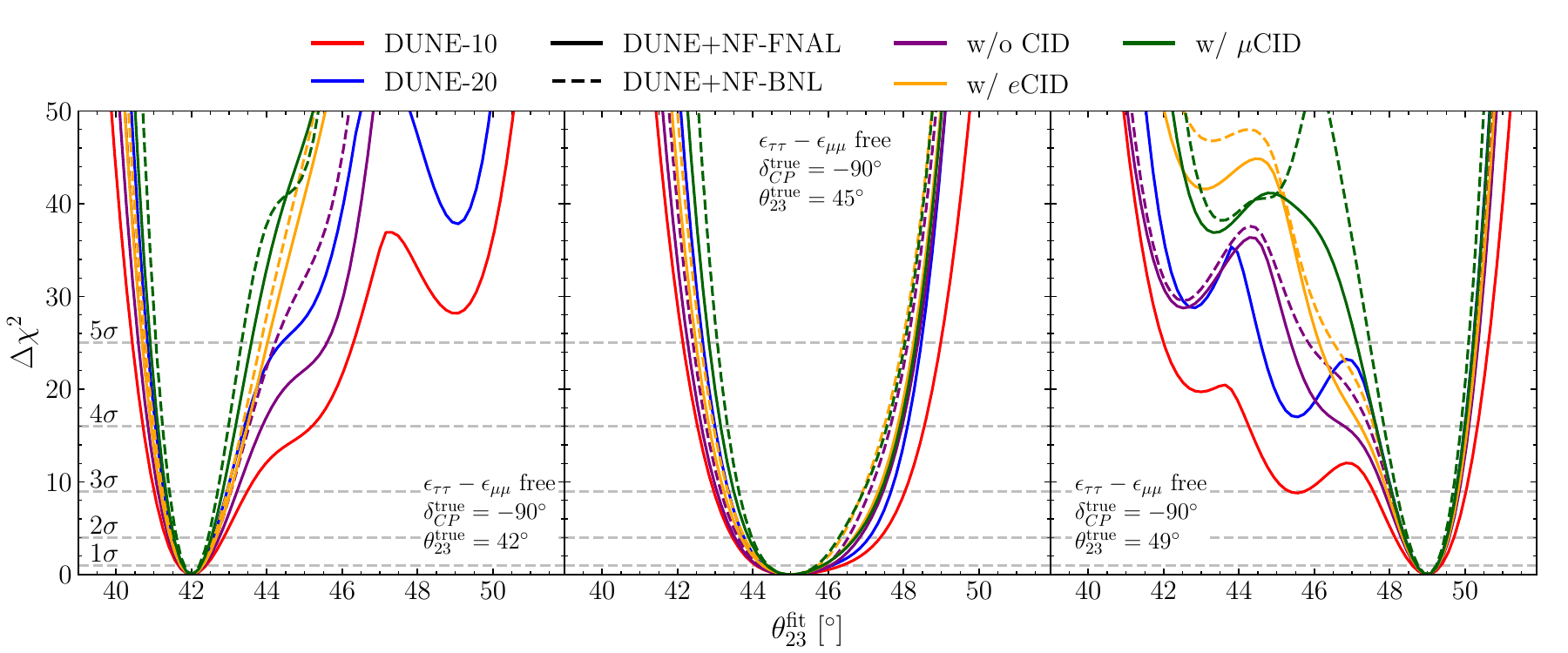}
\caption{The experimental sensitivity to $\theta_{23}$, where the $\epsilon_{\tau\tau}-\epsilon_{\mu\mu}$ is minimized at a time.}
\label{fig:theta23 NSI tt-mm 1D}
\end{figure}

\begin{figure}
\centering
\includegraphics[width=\textwidth]{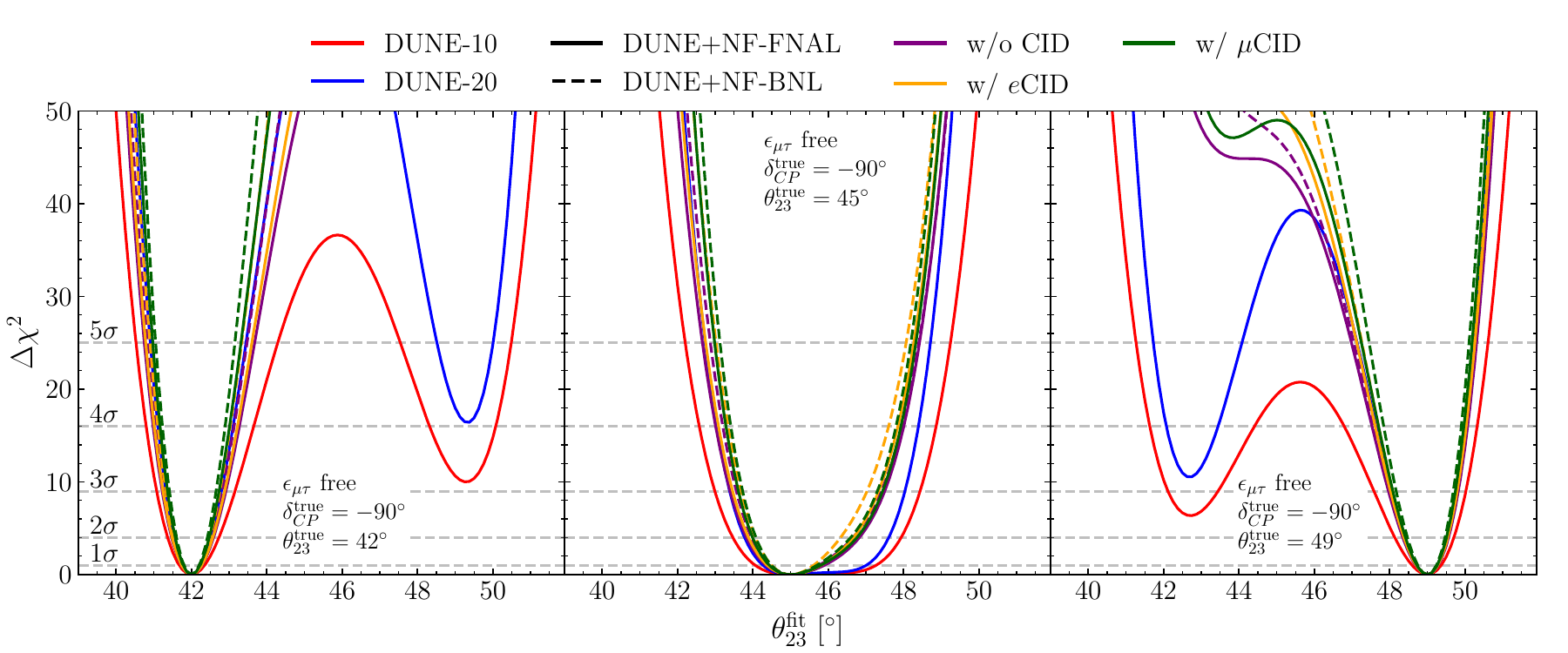}
\caption{The experimental sensitivity to $\theta_{23}$, where the $\epsilon_{\mu\tau}$ is taken as complex and the corresponding phase is minimized together with its modulus at a time.}
\label{fig:theta23 NSI mt 1D}
\end{figure}

\begin{figure}
\centering
\includegraphics[width=\textwidth]{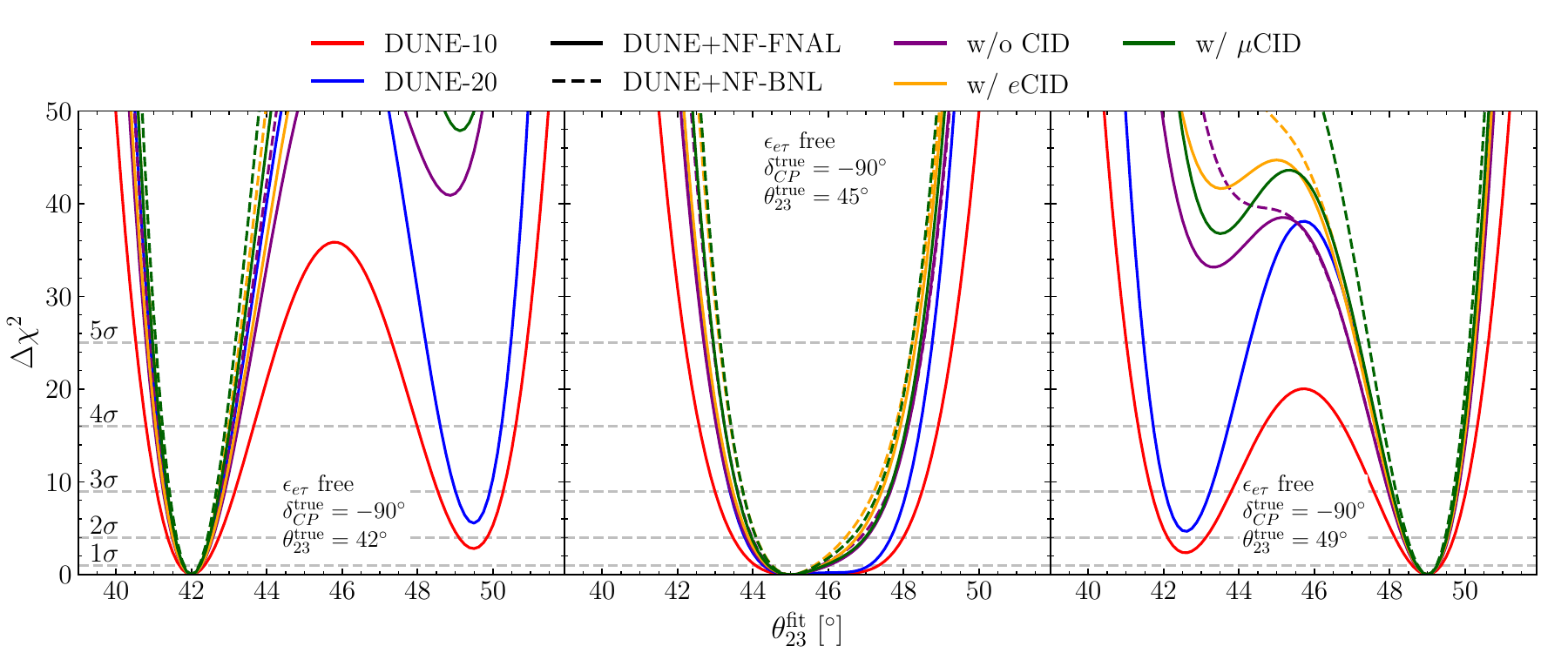}
\caption{The experimental sensitivity to
$\theta_{23}$, where the $\epsilon_{e\tau}$ is 
taken as complex and the corresponding phase is minimized together with its modulus at a time.}
\label{fig:theta23 NSI et 1D}
\end{figure}

In this appendix we present additional results. Figs.~\ref{fig:NSI 1D BNL}, \ref{fig:NSI two-D BNL}, \ref{fig:NSI 2D phase BNL}, \ref{fig:Delta delta NSI 1D minimized BNL}, \ref{fig:CPT 2D BNL}, \ref{fig:CPT 2D zoom BNL}, \ref{fig:CPT 2D benchmark BNL}, \ref{fig:theta23 NSI em 1D BNL} show the results for the BNL setup, similar to figs.~\ref{fig:NSI 1D}, \ref{fig:NSI two-D}, \ref{fig:NSI 2D phase}, \ref{fig:Delta delta NSI 1D minimized}, \ref{fig:CPT 2D}, \ref{fig:CPT 2D zoom}, \ref{fig:CPT 2D benchmark}, \ref{fig:theta23 NSI em 1D} in the main text.
In Fig.~\ref{fig:delta NSI 1D minimized}
we show the precision on $\delta_{CP}$ assuming $\delta_{CP}^{\rm{true}}=0$ in the presence of NSI. Figs.~\ref{fig:theta23 NSI ee-mm 1D}, \ref{fig:theta23 NSI tt-mm 1D}, \ref{fig:theta23 NSI mt 1D}, \ref{fig:theta23 NSI et 1D} we show the sensitivity to $\theta_{23}$ in the presence of non-zero NSI parameters, similar to Fig.~\ref{fig:theta23 NSI em 1D} in the main text.

\bibliographystyle{JHEP}
\bibliography{main}

\end{document}